\documentclass[11pt,twoside,a4paper,openany]{article}
\usepackage[round, sort]{natbib}
\usepackage{setspace}
\usepackage[titletoc,toc]{appendix}
\usepackage{geometry}                
\usepackage[font=small,labelfont=bf]{caption}
\usepackage{hyperref}
\usepackage{xfrac}      
\usepackage[parfill]{parskip}    
\usepackage{graphicx}
\usepackage{amssymb}
\usepackage{epstopdf}
\usepackage{mathtools}
\usepackage{lineno}
\usepackage{rotating}
\usepackage{authblk}
\usepackage{amsmath}
\usepackage{stmaryrd}
\usepackage{fancyhdr}
\usepackage[usenames, dvipsnames]{color}
\author[1]{\small Joshua D. Carmichael}
\affil[1]{\small  Joshua D Carmichael
EES-17, Los Alamos National Laboratory.
Mailing address:
P.O. Box 1663,
MS D446,
Los Alamos, NM 87545.
Email: joshuac@lanl.gov
}
\title{Hypothesis Tests on Rayleigh Wave Radiation Pattern Shapes: A Theoretical Assessment of Idealized Source Screening}
\begin{document}
\onehalfspacing

\date{}
\maketitle
\begin{abstract} 
Shallow seismic sources excite Rayleigh wave ground motion with azimuthally dependent radiation patterns. We place binary hypothesis tests on theoretical models of such radiation patterns to screen cylindrically symmetric sources (like explosions) from non-symmetric sources (like non-vertical dip-slip, or non-VDS faults). These models for data include sources with several unknown parameters, contaminated by Gaussian noise and embedded in a layered half-space. The generalized maximum likelihood ratio tests that we derive from these data models produce screening statistics and decision rules that depend on measured, noisy ground motion at discrete sensor locations. We explicitly quantify how the screening power of these statistics increase with the size of any dip-slip and strike-slip components of the source, relative to noise (faulting signal strength), and how they vary with network geometry. As applications of our theory, we apply these tests to (1) find optimal sensor locations that maximize the probability of screening non-circular radiation patterns, and (2) invert for the largest non-VDS faulting signal that could be mistakenly attributed to an explosion with damage, at a particular attribution probability. Lastly, we quantify how certain errors that are sourced by opening cracks increase screening rate errors. While such theoretical solutions are ideal and require future validation, they remain important in underground explosion monitoring scenarios because they provide fundamental physical limits on the discrimination power of tests that screen explosive from non-VDS faulting sources.
\\ \\
\textbf{Plain Language:} We derive hypothesis tests that compare competing physical models of Rayleigh waves triggered by shallow, buried sources. Our tests sample the Rayleigh wavefield along a circle that is centered at the source to evaluate evidence that data show either (1) a circular radiation pattern or (2) a non-circular radiation pattern, when the focal mechanisms are unknown. The resultant evidence test show an observer's optimal capability to screen most faulting sources from explosion sources depend on sensor number, azimuthal gap, and source focal mechanism. We apply these tests to (1) find the best sensor locations that maximize our ability to screen isotropic sources explosions from most faults, and (2) invert for the largest ``faulting signal'' that an observer could mistakenly attribute to an explosion with damage, at a particular detection rate. Such theoretical solutions reveal physical limits on the screening power of radiation pattern tests that screen explosive from non-vertical dip-slip faulting sources of Rayleigh wave energy.
\end{abstract}
\clearpage
\section{Introduction}\label{sec:intro}
\begin{figure*}
 \centering
\includegraphics[width=1\textwidth]{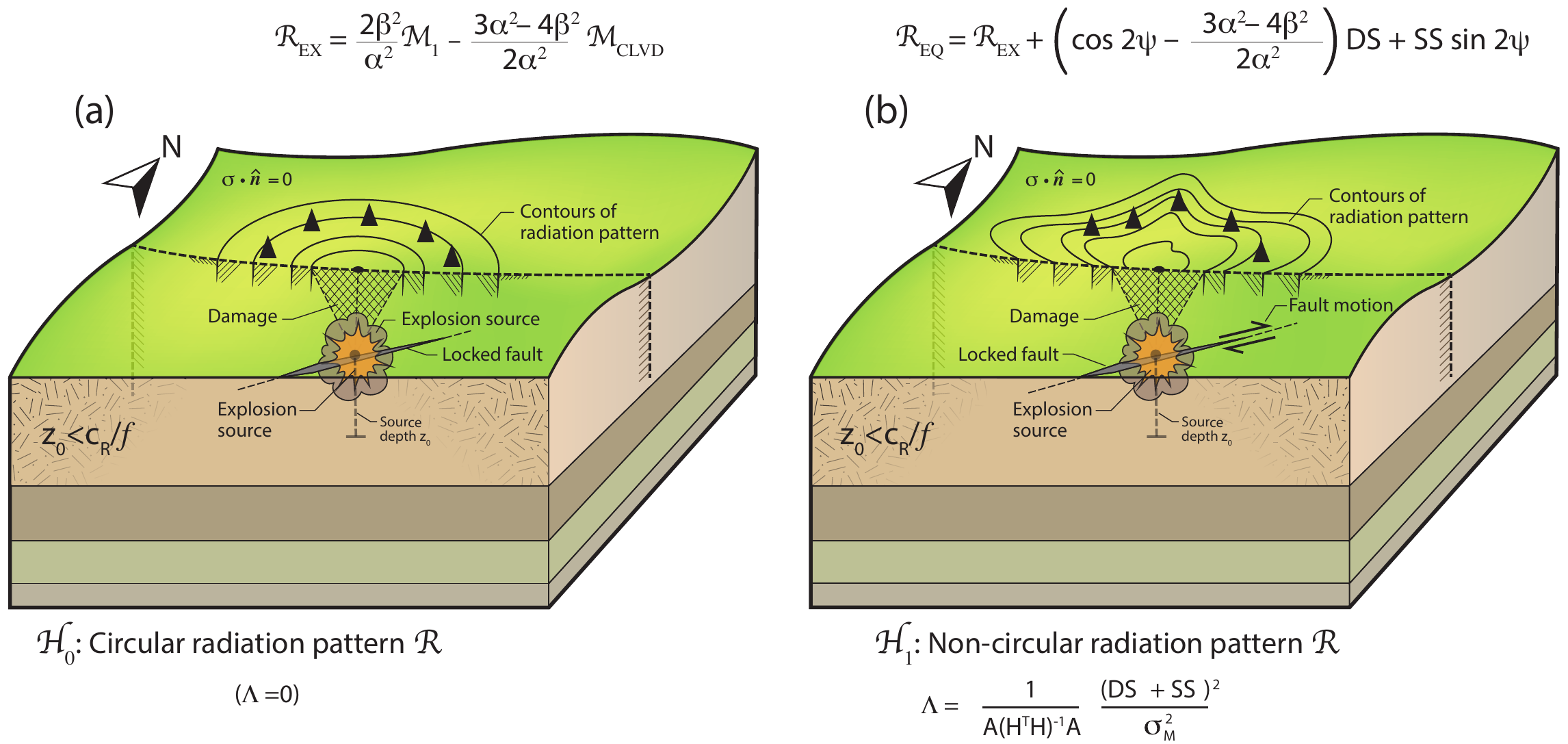}
\captionof{figure}[]
{\narrower Surface sensors (triangles) record Rayleigh waves from a shallow explosion that is buried to a depth $h$ in a horizontally-stratified half-space that has moment $M_{\text{I}}$, which creates damage with moment $M_{\text{CLVD}}$ that each superimpose with faulting of moment $M_{0}$. The damage occurs when constructive interference between the incident and reflected waves create a conically-shaped zone of failure. (\textbf{a}): An explosion source capable of producing damage that is symmetric about a vertical axis, but that includes no faulting (no tectonic release) radiates a circular radiation pattern $\boldsymbol{\mathcal{R}}_{\text{EX}}$ described at top. (\textbf{b}): A source that includes non-VDS faulting near the explosion point radiates a non-circular radiation pattern $\boldsymbol{\mathcal{R}}_{\text{EQ}}$ described at top. Each radiation pattern shape associates with a deterministic scalar $\Lambda$ (bottom) that quantifies the power of the hypothesis test between data that record circular versus non-circular radiation patterns (at top). Subsection \ref{sec:CaseI} defines each symbol.}
 \label{fig:HypTest}
 \end{figure*}

Seismic sources produce Rayleigh wave radiation patterns that indicate focal mechanism asymmetry (Fig. \ref{fig:HypTest}). Teleseismic and regional records of large underground nuclear explosions ($\sim10^{1}-10^{3}$kT) emplaced in pre-stressed geologies, for example, have historically revealed that tectonic release of strain can superimpose with the isotropic explosion source to produce Love and Rayleigh waves with non-circular radiation patterns \citep{Toksoz19721, Ekstrom19941}. These observations contrast with simple models of cylindrically-symmetric, shallow sources hosted in vertically stratified geologies that predict azimuthally constant Rayleigh wave amplitudes and absent Love waves. Instead, such observations are partially consistent with models of faulting motion presented by tectonic release, which produces asymmetric deformation about a vertical axis that excites a Rayleigh-wave field with a non-circular radiation pattern \citep{Richards19951}. Observers risk misidentifying large, shallow explosions that accompany such tectonic release as shallow earthquakes when the imprint of such a ``faulting signal'' on the radiation pattern shape is significantly non-circular or has a low signal-to-noise ratio (SNR). Smaller yield explosions ($\sim10^{1}-10^{3}$kg) that do not trigger tectonic release may still accompany fracturing events or mechanical sliding on rock joints that output shear energy as bulk damage \citep{Steedman20161}. These bulk effects do not appear to significantly distort surface wave radiation patterns when they are symmetric about a vertical axis. In particular, local distance records from a series of conventional explosives conducted through the Source Physics Experiment (SPE) \citep{Snelson20131} demonstrate that low yield, shallowly buried sources that cause damage are vertical-axis symmetric, and thereby trigger Rayleigh wave radiation patterns that are circular within a factor of 1.25  \citep{Larmat20171}. These explosions, which were emplaced in the granites of Climax stock on the Nevada National Security Site (NNSS), also demonstrate that body waves radiated by the same source in a similar frequency band show apparent azimuthally asymmetry that is likely sourced by refraction effects \citep{Darrh20191}. Experiments with low-yield explosives emplaced in boreholes within New Hampshire granite demonstrate more complex effects. Namely, these latter experiments reveal that observationally confirmed radial fracturing in hard rock accompanies asymmetric, short period ($\sim$6Hz), guided Rayleigh waves (Rg). Body waves sourced by these same New Hampshire explosions can exhibit even greater azimuthal dependence \citep{Stroujkova20181}. 
%
Both the SPE and New Hampshire experiments cumulatively indicate two substantive effects of the seismic source on radiation pattern shape can be observed from local distances. First, vertical axis-symmetric damage does not significantly distort Rayleigh wave radiation pattern shapes from circularity, whereas a structure at depth can distort body wave patterns. Second, significant radial fracturing and shearing motion at the shallow explosion source measurably distorts both Rayleigh wave and body wave radiation patterns. These observations indicate that azimuthal variability of Rayleigh waves sourced by small yield explosions indicate faulting or shearing motion at the source, whereas body wave asymmetry appears non-uniquely attributed to source and path.

A forthcoming test within the SPE series will provide a unique opportunity to study asymmetry in radiation patterns that are sourced by explosions and superimposed with historic, tectonic records of fault slip. This effort will detonate a conventional explosive in tectonic regions and monitor its seismic signatures for comparison against earlier earthquake signatures that output radiation that shares wavefront paths \citep{Walter20121}. If Rayleigh waves output by this explosion remains circular and mismatches historical records, theory and the aforementioned observations imply that the resultant radiation pattern shape provides a hopeful discriminant. It remains unclear, however, ``how much'' non-circularity of the Rayleigh wave radiation pattern is sufficient to indicate that a seismic source is dominated by shearing or fault motion, and is therefore more earthquake-like than explosion-like. A discriminant that tests radiation pattern distortion from local to teleseismic ranges can therefore benefit both the source physics community and support nuclear test-ban treaty surveillance, if justified on both theoretical and experimental grounds.

This work determines if a discriminant for shallow source types that tests Rayleigh wave radiation pattern shapes is justified on theoretical grounds. To confront this challenge, we apply binary hypothesis tests to competing data models of noisy radiation patterns and assume several source parameters are unknown. Our models also use planar geologies and therefore isotropic propagation of Rayleigh waves; we do not consider Love waves. Under this limited scope, we derive screening curves from the hypothesis tests that bound an observers's capability to discriminate non-vertical dip-slip (non-VDS) shallow faulting sources from symmetric sources (like shallow explosions that create damage that is symmetric about a vertical axis). The screening power of these tests is completely defined by the product of a faulting-signal term, and a term that depends on deployment parameters. Using this test, we illustrate several applications of our theory that include finding deployment locations for sensors to supplement an existing network and that maximize our source-type screening probability. We conclude that a Rayleigh wave radiation pattern discriminant shows a high probability $\text{Pr}_{D}$ of success ($\text{Pr}_{D}$ $>$ $0.9$) when a sufficient number of sensors ($>$ $12$) with a limited azimuthal gap ($<90^{\circ}$) records radiation from sources with moderate strike-slip faulting signals (SNR $> 20$). This probability diminishes with more ambiguous focal mechanisms that resemble certain historical events recorded near the Lop Nor nuclear test site in China. Our results suggest that a screening statistic that tests Rayleigh wave radiation pattern shapes for faulting signals requires very dense sensor coverage to achieve good efficacy, and is probably impractical in most passive monitoring scenarios. We lastly consider how radiation pattern distortion from opening cracks and wavefield homogenization can inflate screening errors.

We emphasize that our treatment is purely theoretical and idealized. Our models require that future efforts provide any evidence that this approach has a practical utility as a screening technique in non-ideal settings. In particular, focusing effects \citep{Selby20001, Lin20121}, tectonic pre-stress \citep{Toksoz19711, Ekstrom19941}, surface roughness \citep{Steg19701}, path-dependent attenuation \citep{Macbeth19871}, transmission losses through vertical interfaces \citep{Napoli20181}, and scattering from topography that is comparable in relief to Rayleigh wavelengths \citep{Snieder19861,Ichinose20171} often dominate the azimuthal variability of observed surface wave radiation. Stratified half-space models that support idealized Rayleigh wave propagation are likely limited to very local seismic measurements of SPE shots, cyroseismic studies interior to crevasse-free ice sheets and glaciers \citep{Lindner20191, Carmichael20151, Lough20151, Hudson20201}, and to engineered structures \citep{Lu20071}. We assert that our study remains useful in geophysics, however, because it helps define the physical limits on an observer's capability to screen Rayleigh wave sources, using only their radiation pattern shapes. Such limits support previous efforts that exploit Rayleigh wave radiation pattern geometry to better characterize underground explosions \citep{Toksoz19711, Yacoub19811}, and supports present computational and observational work to understand earthquake radiation \citep{Rosler20201}.
%
\begin{figure*}
 \centering
\includegraphics[width=1\textwidth]{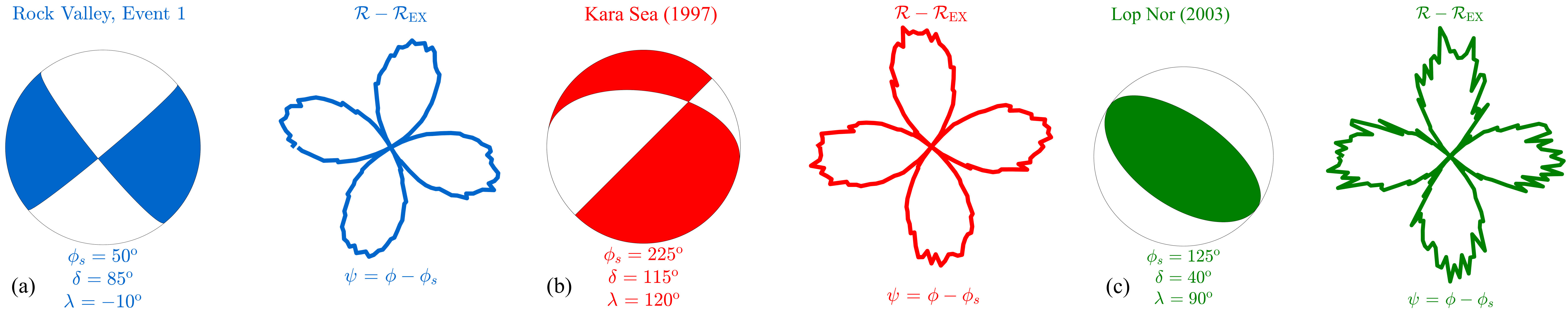}
\captionof{figure}[]
{\narrower Three focal mechanism solutions for shallow earthquakes that locate near underground test sites and the noisy, non-circular component of the radiation pattern for each source. (\textbf{a}): The $M_{W}$ $=$ $3.7$, 1993-05-30 event from the Rock Valley Sequence, termed ``Event 1'' (depth $\approx$ 2 km) \citep{Smith19931}. (\textbf{b}): The $M_{L}$ $=$ $3.3$, 1997-08-16 event from the Kara Sea (0 km $<$ depth $<$ 23 km) that located tens of km from Novaya Zemlya \citep{Hartse19981,Schweitzer20071,Bowers20021}. \textbf{c}: The $m_{b}$ $=$ $4.8$, 2003-03-13 tectonic event located near the Lop Nor test site  (5 km $<$ depth $<$ 7 km) \citep{Selby20051}. In each pattern, Gaussian noise superimposes with a deterministic radiation pattern solution to mimic a finely sampled observation with an overall SNR of 20. We note that the noise on the non-circular component of the radiation pattern increases from left to right, as the sources produce Rayleigh waves less efficiently. We further assume that the isotropic component of the source is zero in each case, and within the bound of any uncertainties in the focal mechanism estimate. Here, $M_{W}$ is the moment magnitude, $M_{L}$ is local or Richter magnitude, and $m_{b}$ is body wave magnitude.}
 \label{fig:allRadPatts}
 \end{figure*}
 \section{Reference Sources and Historical Events}
 Our assessment will compare cylindrically symmetric sources that include explosions with damage along the vertical axis against a set of four distinct non-VDS, tectonic earthquake sources. One such faulting source is the strike-slip earthquake. This particular double-couple source is maximally dissimilar from any isotropic source, as plotted on a Hudson diagram \citep{Hudson19891} or lune \citep{Tape20191}. We also consider historical, tectonically sourced events that are shallow and locate near former underground testing sites. These events include the Rock Valley fault  located within the Nevada National Security Site complex in the US, a regionally recorded seismic event beneath the Kara Sea near Novaya Zemlya, and a shallow source located near the Chinese Lop Nor nuclear test site. Fig. \ref{fig:allRadPatts} pairs the beach ball representation of the focal mechanisms for these sources with noisy representations for the non-circular part of their radiation patterns. The confidence region for the hypocentral depth of each observed source includes depths shallow enough to be reached by current drilling capabilities (within 7km of Earth's surface). We shall often refer to such shallow, non-VDS tectonic earthquakes as simply non-VDS faulting sources. Similarly, we shall refer to explosions that may or may not accompany damage that is symmetric about a vertical axis as just an explosive source. Any such damage preserves the circularity of Rayleigh wave radiation patterns.
\section{Shallow Source Rayleigh Waves} \label{sec:ShallSourceTheory}
We consider Rayleigh wave motion present in a cylindrically symmetric, horizontally layered half-space triggered by a buried seismic source (Fig. \ref{fig:HypTest}). This source may be described by a point force or a symmetric moment $3\times3$ moment tensor $\boldsymbol{M}$ that summarizes displacement discontinuities as sets of force couples. Such a moment tensor source excites the frequency-domain, Rayleigh wave displacement $\boldsymbol{u}$ $=$ $[u_x$, $0$, $u_{z}]^{\text{T}}$ that solves Newton's second law of motion in this stratified half-space, subject to traction-free surface boundary conditions. The scalar product of $\boldsymbol{M}$ with the gradient of the (Hermitian) Rayleigh wave Green's tensor $\boldsymbol{G}^{\text{RAY}}$ \cite[eq. 3.23]{Aki20021} compactly represents such a general displacement: 
\begin{equation}
\label{eq:gendispl}
\boldsymbol{u} = \boldsymbol{M} \cdot \nabla  \boldsymbol{G}^{\text{RAY}}. 
\end{equation}
When the source is emplaced in half-space with non-trivial stratification (not pure half-space), the displacement solution of eq. \ref{eq:gendispl} is a linear combination of eigenvectors $\boldsymbol{r}$ that sum over distinct wavenumber modes \cite[eq. 7.150-7.151]{Aki20021}. The dispersion relationship $k$ $=$ $k\left( \omega \right)$ for this Rayleigh wave field defines these modes and couples the radiation pattern $\mathcal{R}$ of $\boldsymbol{u}$ to the range- and depth- dependent part of the displacement solution $\boldsymbol{g}$. If the moment tensor source is shallow enough that the free-surface traction free boundary conditions $\boldsymbol{\sigma} \cdot \hat{\boldsymbol{e}_{z}}$ $=$ $\boldsymbol{0}$ also apply at the source's depth, then the radiation pattern effectively decouples from $\boldsymbol{g}$ \cite[pg. 328]{Aki20021}; here, $\boldsymbol{\sigma}$ is the elastic stress tensor, $\hat{\boldsymbol{e}_{z}}$ is a unit vector pointing vertically upward, and $\boldsymbol{0}$ is a three element vector of zeros. More formally, this condition requires that the source is buried at a depth that is small compared to the dominant wavelength of the seismic radiation triggered by the source-time function. These cumulative conditions mean the Rayleigh wave field has the simple form and is a product of $\mathcal{R}$ and $\boldsymbol{g}$:
\begin{equation}
\begin{split}
\label{eq:displ}
\boldsymbol{u} &= \mathcal{R}\left(  \phi \right) \cdot \boldsymbol{g}( \boldsymbol{x}, \omega )
\\
&= \left[ \bar{\mathcal{R}} + \mathcal{R}_{1} \cos \left( 2 \phi \right) + \mathcal{R}_{2} \sin \left( 2 \phi \right) \right] \cdot  \boldsymbol{g}( \boldsymbol{x}, \omega ).
\end{split}
\end{equation}
Vector $\boldsymbol{g}$ in eq. \ref{eq:displ} depends on source depth and observation range, but not $\phi$; term $\bar{\mathcal{R}}$ is the mean radiation pattern; $R_m$ ($m = 1,2$) are non-circular radiation pattern coefficients that all have units of moment and depend on moment tensor $\boldsymbol{M}$; and $\phi$ is the azimuthal angle (see Aki and Richards, 2002, Fig. 4.20). The general displacement in eq. \ref{eq:gendispl} and the traction free boundary conditions relate the mean radiation pattern and azimuthally dependent coefficients to linear functions of the moment tensor elements:
\begin{equation}
\label{eq:cartRadPatt}
\begin{split}
\bar{\mathcal{R}} &= \frac{1}{2} \left( M_{xx} + M_{yy}\right) - \left(1 - \frac{2\beta^{2}}{\alpha^{2}}\right) M_{zz}
\\
\mathcal{R}_{1} &= \frac{1}{2} \left( M_{xx} - M_{yy}\right)
\\
\mathcal{R}_{2} &= M_{xy}
\end{split}
\end{equation}
where $M_{ij}$ is the force couple aligned with the Cartesian direction $i$ separated by direction $j$. We note that Rayleigh wave radiation patterns do not resolve terms $M_{xz}$ and $M_{yz}$ that indicate vertical shearing motions directed parallel to Earth's surface.
\subsection{Colocated Faulting, Explosion and Damage Sources}
The most general symmetric moment tensor for a single point source is a superposition of an explosion with isotropic moment $M_I$ that effectively colocates with a non-VDS faulting source of moment $M_0$ and a damage source, which is equivalent to a compensated linear vector dipole (CLVD) with moment $M_{\text{CLVD}}$. Moment $M_{\text{CLVD}}$ is positive if explosions create dilatation along the vertical axis of damage, and negative if explosions drive contraction along that axis. We assume that the fault is parameterized by strike ($\phi_{s}$), rake ($\lambda$) and dip ($\delta$) angles. The radiation pattern coefficients from eq. \ref{eq:displ} are then (algebra omitted):
\begin{equation}
\label{eq:damageRadPatCoeffs}
\begin{split}
\bar{\mathcal{R}} &= \cfrac{2 \beta^{2}}{\alpha^{2}}M_{I} - \cfrac{3\alpha^{2} - 4\beta^{2}}{2 \alpha^{2}} M_{\text{CLVD}}  - \cfrac{3\alpha^{2} - 4\beta^{2}}{\alpha^{2}} DS  \\
\mathcal{R}_{1} &= DS \cos \left( \phi_{s} \right) - SS \sin \left( \phi_{s} \right) \\
\mathcal{R}_{2} &=   SS \cos \left( \phi_{s} \right) + DS \sin \left( \phi_{s} \right)
\end{split}
\end{equation}
where we have used a moment decomposition \cite[eq. 15]{Patton20111}. In eq. \ref{eq:damageRadPatCoeffs}, scalar $SS$ quantifies the strength of the strike-slip faulting component, whereas scalar $DS$ quantifies the strength of the dip-slip component. These terms are \cite[eq. 21 and eq. 22]{Ekstrom19941}:
\begin{equation}
\label{eq:tectRelease}
\begin{split}
DS &= \cfrac{1}{2} M_{0} \sin\left(2 \delta \right) \sin\left(\lambda \right) \\
SS &= M_{0} \sin\left( \delta \right) \cos\left(\lambda \right)
\end{split}
\end{equation}
where the equivalent expression for $SS$ elsewhere \cite[eq. 16]{Patton20111} is missing a factor of two in the dip angle. We express the full Rayleigh wave displacement by combining eq. \ref{eq:displ}, eq. \ref{eq:damageRadPatCoeffs}, and eq. \ref{eq:tectRelease}. The superposition of the explosive, damage and non-VDS faulting sources thereby creates a Rayleigh wave displacement field that is:
\begin{equation}
\label{eq:displFaultWithDamage}
\boldsymbol{u} =  \left[ \bar{\mathcal{R}}  + DS \cos( 2  \psi   ) +  SS \sin( 2  \psi   ) \right] \cdot \boldsymbol{g}
\end{equation}
where $\psi$ $=$ $\phi$ $-$ $\phi_{s}$ is the difference between the azimuthal and strike angles; because we use $\psi$ hereon, we refer to it as the ``azimuthal'' angle, despite abuse in terminology. Regardless, the function that scales $\boldsymbol{g}$ is the radiation pattern $\mathcal{R}$, and includes the same functional form as the scalar factor that appears in eq. \ref{eq:displ}:
\begin{equation}
\label{eq:tecRadPat}
\begin{split}
\mathcal{R} &=  \bar{\mathcal{R}} + DS \cos( 2 \psi )  +  SS \sin( 2 \psi )
\end{split}
\end{equation}
Eq. \ref{eq:damageRadPatCoeffs} with eq. \ref{eq:tecRadPat} demonstrates that $ \bar{\mathcal{R}} $ is azimuthally constant, and that azimuthal variability arises from non-VDS faults where $\delta$ $\neq$ $\pm\pi$/2, $\lambda$ $\neq$ $\pm\pi$/2. Eq. \ref{eq:tecRadPat} is also the starting point for estimating the optimal receiver sampling of the Rayleigh wave radiation pattern of shallow sources (Section  \ref{sec:OptimalDeployment}).
\section{Statistical Hypothesis Tests} \label{sec:hypotTest}
To determine if observed Rayleigh wave radiation indicates an isotropic or mixed non-VDS source, we evaluate a hypothesis test that compares two distinct models for deterministic signals $\mathcal{R}$. This test forms a generalized likelihood ratio statistic from $N$ observations of $\mathcal{R}$ at different azimuthal locations and compares them to a threshold to measure evidence that the data record a non-circular radiation pattern. Specifically, our null hypothesis is that an azimuthally distributed set of sensors samples a circular radiation pattern with zero non-VDS faulting moment ($M_{0} =\, 0$). Our alternative hypothesis states that these same sensors sample a non-circular radiation pattern. The competing hypotheses are (without noise):
\begin{equation}
\begin{split}
\mathcal{H}_0:  \mathcal{R} \bigr\vert_{M_{0} =\, 0}&=  \bar{\mathcal{R}}\bigr\vert_{M_{0} =\, 0}
\\
\mathcal{H}_1:  \mathcal{R}  \bigr\vert_{M_{0} >\, 0} &= \bar{\mathcal{R}}\bigr\vert_{M_{0} =\, 0} + DS \cos( 2 \psi )  +  SS \sin( 2 \psi )
\end{split}
\label{eq:HypotTest}
\end{equation}
where $\delta$, $\lambda$, and $M_{0}$ parameterize $DS$ and $SS$, and are unknown. We now add a random variability models to eq. \ref{eq:HypotTest} in two respects: first, we superimpose Gaussian measurement noise to $\mathcal{R}$, and second, we model $\mathcal{R}$ itself to be a sum of the deterministic component and a zero mean random process. To include measurement noise, we note that observations often show that moment estimates are log-normally distributed if such moment is measured from station magnitudes \citep{Godano20001,Kagan20021}. Moment estimates that fit a horizontal line to the low-frequency component of waveform's displacement spectrum reveal Gaussian residuals, because normally distributed errors dominate seismic waveform records \citep{Vsileny19961,Walter20091}. We therefore assume waveform noise is the dominant source of random error in scalar moment estimates and model a single observation at fixed azimuth as a Gaussian random variable. To modify the signal portion of the radiation pattern, we add a zero-mean Gaussian process with covariance $\sigma_{\mathcal{R}}^{2}$ to the deterministic, non-zero mean of $\mathcal{R}$. We indicate that $\mathcal{R}$ is a random variable with mean $\mu$ and variance $\sigma^{2}$ with notation $\mathcal{R}$ $\sim$ $\mathcal{N}\left(\mu, \sigma^{2} \right)$, where $\sim$ indicates ``is distributed as''. Eq. \ref{eq:HypotTest} then tests between two competing probability distributions for $\mathcal{R}$ (with noise):
\begin{equation}
\begin{split}
\mathcal{H}_0: \,\,   \mathcal{R} &\sim \mathcal{N}\left( \mathcal{R} \bigr\vert_{M_{0} =\, 0}, \,\, \sigma_{M}^{2} \right)
\\ 
\mathcal{H}_1: \,\,  \mathcal{R} &\sim  \mathcal{N}\left( \mathcal{R} \bigr\vert_{M_{0}\,>\,0}, \,\, \sigma_{M}^{2} + \sigma_{\mathcal{R}}^{2} \right).
\end{split}
\label{eq:distrHypotTest}
\end{equation}
Eq. \ref{eq:distrHypotTest} applies to one observation $\psi$. For many azimuthally distributed observations, our hypothesis test  includes a vectorized set of radiation pattern samples $\boldsymbol{\mathcal{R}}$ where the $k^{\text{th}}$  sample is $ \mathcal{R}_{k}$:
\begin{equation}
 \boldsymbol{\mathcal{R}} \triangleq \left[ \mathcal{R}_{1}, \, \mathcal{R}_{2}, \,\cdots, \mathcal{R}_{k}, \, \cdots, \mathcal{R}_{N}\right]^{\text{T}}.
 \label{eq:vecRadPat}
\end{equation}
The competing hypotheses (eq. \ref{eq:distrHypotTest}) form a general Gaussian detection problem, applied to a multi-sample radiation pattern \cite[section 5.6]{Kay19981}. To explicitly define each radiation pattern term, we concatenate the unknown statistical parameters into a single vector. We then express the mean $\boldsymbol{\mu}$ of $\boldsymbol{\mathcal{R}}$ under $\mathcal{H}_1$ as a linear system that equates with eq. \ref{eq:damageRadPatCoeffs} so that component $m$ (row $m$) of $\boldsymbol{\mu}$  is:
\begin{equation}
\mathcal{\mu}_{m} = 
\underbrace{\begin{bmatrix}
1 &  \left( \cos( 2 \psi_{m} ) - c \right)  &  \sin( 2 \psi_{m} )
\end{bmatrix}}_{\boldsymbol{H}(m,:)}
\cdot
\underbrace{
\begin{bmatrix}
\Delta  \bar{\mathcal{R}} 
 \\ 
 DS 
 \\
  SS
\end{bmatrix}
}_{\boldsymbol{\theta}}
\label{eq:systemMatrix}
\end{equation}
where: 
\begin{equation}
\begin{split}
\Delta  \bar{\mathcal{R}}  &\triangleq  \bar{\mathcal{R}}  + c DS =  \cfrac{2 \beta^{2}}{\alpha^{2}}M_{I} - \cfrac{3\alpha^{2} - 4\beta^{2}}{2 \alpha^{2}} M_{\text{CLVD}} 
\\
c &\triangleq \cfrac{3\alpha^{2} - 4\beta^{2}}{\alpha^{2}}
\end{split}
\label{eq:defineTheta}
\end{equation}
and where $\boldsymbol{H}(m,:)$ in eq. \ref{eq:systemMatrix} defines row $m$ of matrix $\boldsymbol{H}$. The first parameter $\Delta  \bar{\mathcal{R}} $ weights the circular portion of the radiation field whereas the other parameters weight the non-circular portion. The covariance between the observations $\mathcal{R}_{i}$ and $\mathcal{R}_{j}$ at two azimuths is zero (eq. \ref{eq:covRzero}). This independence in observations implies that the probability density function (PDF) for vector $ \boldsymbol{\mathcal{R}}$ under  $\mathcal{H}_1$ has mean $\boldsymbol{\mu}$  $=$ $\boldsymbol{H} \boldsymbol{\theta}$ and covariance $(\sigma_{\mathcal{R}}^{2}+\sigma_{M}^{2})\boldsymbol{I}$. The PDF under $\mathcal{H}_{1}$ is therefore:
\begin{equation}
\label{eq:altPdf}
\begin{split}
f_{ \boldsymbol{\mathcal{R}}}\left(  \boldsymbol{\mathcal{R}}; \mathcal{H}_{1} \right) 
&= \frac{1} {\left( 2 \pi (\sigma_{\mathcal{R}}^{2}+\sigma_{M}^{2}) \right)^{\frac{N}{2}}} \text{exp} \left[- \frac{ \vert \vert  \boldsymbol{\mathcal{R}} - \boldsymbol{H} \boldsymbol{\theta}  \vert \vert^{2} }
 {2 \left( \sigma_{\mathcal{R}}^{2}+\sigma_{M}^{2}  \right) } \right]
\end{split}
\end{equation}
where $\boldsymbol{I}$ is the identity matrix.
Under $\mathcal{H}_{0}$, $DS$ $=$ $SS$ $=$ $\sigma_{\mathcal{R}}^{2}$ $=$ $0$. The mean $\boldsymbol{\mu}$ of $\boldsymbol{\mathcal{R}}$ under $\mathcal{H}_0$ is also linear system, but $\boldsymbol{\theta}$ is now constrained to make $\boldsymbol{\mathcal{R}}$ circular:
\begin{equation}
\label{eq:nullPdf}
\mathcal{H}_0: \boldsymbol{A}\boldsymbol{\theta} = \boldsymbol{0}, \quad \text{where:} \quad \boldsymbol{A} = \begin{bmatrix}0 & 1 & 1\end{bmatrix}.
\end{equation}
The PDF for vector $ \boldsymbol{\mathcal{R}}$ with covariance $\sigma_{M}^{2}\boldsymbol{I}$ under $\mathcal{H}_{0}$ is then:
\begin{equation}
\label{eq:nullPdf0}
\begin{split}
f_{ \boldsymbol{\mathcal{R}}}\left(  \boldsymbol{\mathcal{R}}; \mathcal{H}_{0} \right) 
&= \frac{1} {\left( 2 \pi \sigma_{M}^{2} \right)^{\frac{N}{2}}} \text{exp} \left[- \frac{ \vert \vert  \boldsymbol{\mathcal{R}} - \boldsymbol{H} \boldsymbol{\theta}  \vert \vert^{2} }
 {2 \sigma_{M}^{2} } \right] \biggr\vert_{\boldsymbol{A}\boldsymbol{\theta} = \boldsymbol{0}}
\end{split}
\end{equation}
These competing PDFs provide the basis for a binary hypothesis test between circular versus noncircular radiation patterns. The hypothesis test in eq. \ref{eq:HypotTest} that uses such vectorized data is:
\begin{equation}
\begin{split}
\mathcal{H}_0: \,\,   \boldsymbol{\mathcal{R}} &\sim \mathcal{N}\left( \boldsymbol{H} \boldsymbol{\theta}, \,\, \sigma_{M}^{2}\boldsymbol{I} \right), \,\, \boldsymbol{A}\boldsymbol{\theta} = \boldsymbol{0}
\\ 
\mathcal{H}_1: \,\,   \boldsymbol{\mathcal{R}} &\sim  \mathcal{N}\left(\boldsymbol{H} \boldsymbol{\theta}, \,\, \left(\sigma_{M}^{2} + \sigma_{\mathcal{R}}^{2} \right)\boldsymbol{I}  \right).
\end{split}
\label{eq:HypotTest-vec}
\end{equation}
To exploit these tests, we consider two cases. This first case (Case I) considers the variability $\sigma_{\mathcal{R}}$ in $\mathcal{R}$ from random processes as known, but the stochastic variability $\sigma_{M}$ as unknown. The second case (Case II) considers $\sigma_{M}$ as approximately known, but $\sigma_{\mathcal{R}}$ as unknown, though in much less detail than Case I. Parameter $\boldsymbol{\theta}$ is unknown in both cases.  Case I and Case II both include the scenario that $\sigma_{\mathcal{R}}$ is zero as a special example.
%
\begin{figure}
 \centering
\includegraphics[width=\textwidth]{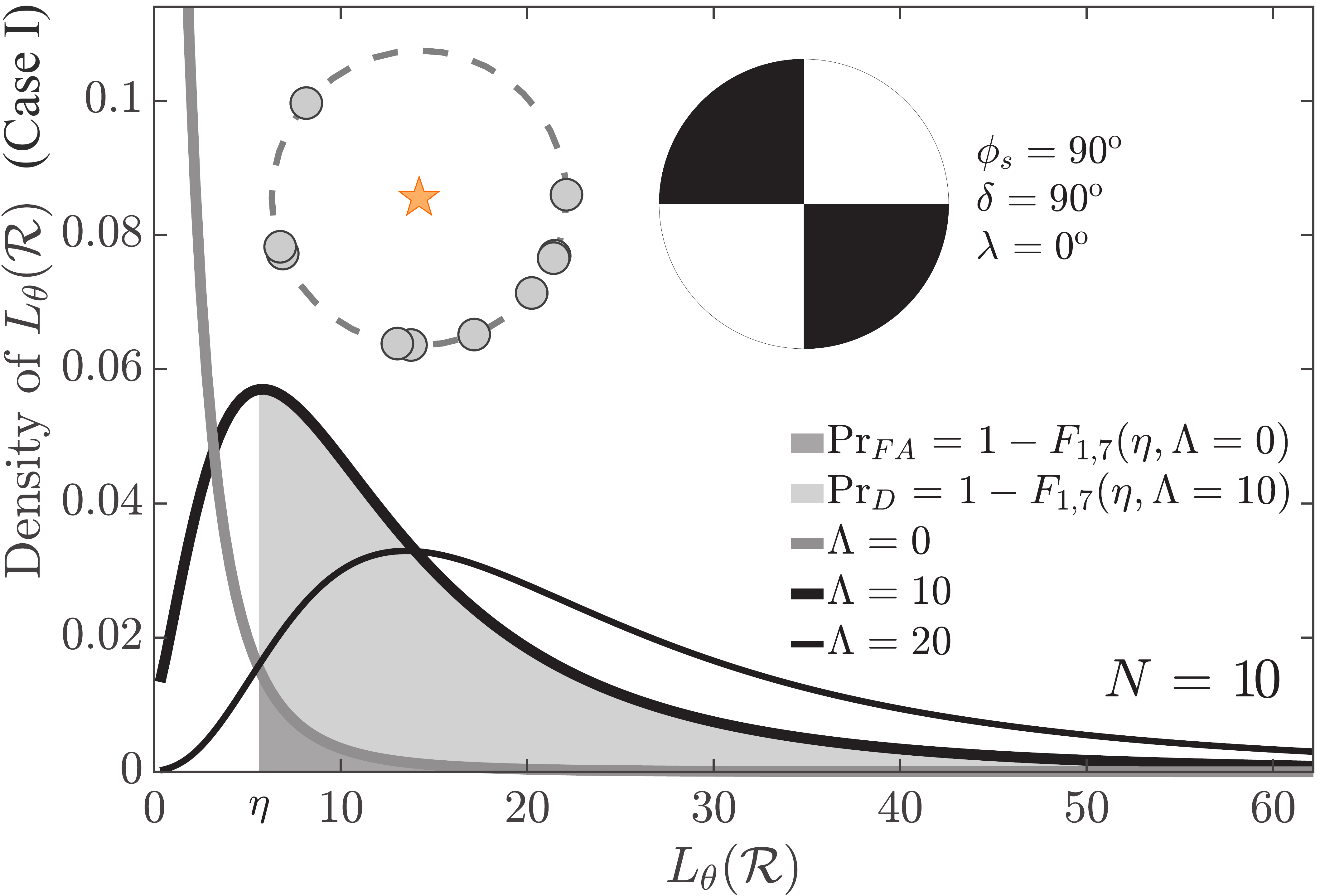}
\captionof{figure}[]
{\narrower Probability density functions that describe the source screening statistic $L_{\boldsymbol{\theta}}\left( \boldsymbol{\mathcal{R}} \right)$ (horizontal axis) in eq. \ref{eq:testStat1}. The statistic describes $N$ $=$ $10$ receivers randomly distributed over a circle centered at the hypocenter of a shallow seismic source (orange star) that sample a source radiation pattern $\boldsymbol{\mathcal{R}}$. The leftmost thick, gray curve indicates a circular radiation pattern ($\Lambda$ $=$ $0$) and rightmost curves indicate noncircular radiation patterns ($\Lambda$ $>$ $0$). The darkest shaded area under the null PDF curve ($\Lambda=0$) indicates a $\text{Pr}_{FA}$ $=$ $5\cdot10^{-3}$ false screening probability. The lighter shaded area below the alternative PDF ($\Lambda=10$) measures the screening probability $\text{Pr}_{D}$ $=$ $0.81$. The threshold is $\eta$ $\approx$ $5$ (eq. \ref{eq:PrFA}).}
 \label{fig:ncfpdf}
 \end{figure}
\subsection{Case I: $\sigma_{\mathcal{R}}$ known, $\boldsymbol{\theta}$ and $\sigma_{M}$ unknown} \label{sec:CaseI}
We first assume a network of $N\ge8$ sensors records a Rayleigh wave radiation pattern with known process variance ($\sigma_{\mathcal{R}}^{2}$ in eq. \ref{eq:HypotTest-vec}), unknown faulting parameters ($\boldsymbol{\theta}$ in eq. \ref{eq:systemMatrix}), and unknown stochastic variance ($\sigma_{M}^{2}$ in eq. \ref{eq:HypotTest-vec}). This case idealizes explosion monitoring scenarios in which an underground testing complex colocates with a shallow fault of unknown geometry. This case also includes the scenario that all variability is stochastic in origin ($\sigma_{\mathcal{R}}^{2}$ $=$ $0$) as an example. Under this general model, a sensor network then records a low magnitude seismic event that originates near a test site, and an observer must estimate the probability that this event is explosive or tectonically sourced by a non-VDS fault (hypothesis $\mathcal{H}_{0}$ or $\mathcal{H}_{1}$).

To determine which hypothesis Rayleigh wave records likely describe, we use a generalized log-likelihood ratio test (log-GLRT). This test compares the logarithmic ratio of the alternative hypothesis PDF to the null hypothesis PDF, where each density is evaluated at maximum likelihood estimates (MLE) of its unknown parameters to form a test statistic. The log-GLRT is equivalent to a conditional decision rule on this scalar test statistic $L_{\boldsymbol{\theta}}^{(I)}\left( \boldsymbol{\mathcal{R}} \right)$:
\begin{equation}
\begin{split}
L_{\boldsymbol{\theta}}^{(I)}\left( \boldsymbol{\mathcal{R}} \right)
\,
&\underset{\mathcal{H}_{0}}
{ \overset{\mathcal{H}_{1}}
{\gtrless}}
\,
\eta \quad \text{where:}
\\ 
L_{\boldsymbol{\theta}}^{(I)}\left( \boldsymbol{\mathcal{R}} \right) &= \ln \left[ \cfrac{ \displaystyle \max_{ {\boldsymbol{\theta},\, \sigma_{M}^{2}}  } \{ \, f_{\boldsymbol{\mathcal{R}}} \left(\boldsymbol{\mathcal{R}}; \mathcal{H}_{1} \right) \, \} }{\,  \displaystyle \max_{ {\boldsymbol{\theta},\, \sigma_{M}^{2}}  } \{ f_{\boldsymbol{\mathcal{R}}} \left( \boldsymbol{\mathcal{R}}; \mathcal{H}_{0} \right) \, \} } \right]
\end{split}
\label{eq:maxLikeGLR}
\end{equation}
We compute the MLEs of the parameters in Appendix \ref{app:CaseI} using projector matrices and reduce the scalar statistic into a ratio of two scaled, statistically independent terms. This provides a test statistic $L_{\boldsymbol{\theta}}\left( \boldsymbol{\mathcal{R}} \right)$ equivalent to eq. \ref{eq:maxLikeGLR} for screening circular from non-circular radiation patterns:
\begin{equation}
L_{\boldsymbol{\theta}}\left( \boldsymbol{\mathcal{R}} \right) = (N-3)\cfrac{\sigma_{M}^{2}}{\sigma_{M}^{2} + \sigma_{\mathcal{R}}^{2}}  \cdot \cfrac{ \bigr \Vert \boldsymbol{P}_{\boldsymbol{X}}\boldsymbol{\mathcal{R}} \bigr \Vert^{2} } { \bigr \Vert \boldsymbol{P}_{\boldsymbol{H}}^{\perp} \boldsymbol{\mathcal{R}} \bigr \Vert^{2}} \underset{\mathcal{H}_{0}}{ \overset{\mathcal{H}_{1}}
{\gtrless}}
\,
\eta.
\label{eq:testStat1}
\end{equation}
The rank-1 projector matrix  $\boldsymbol{P}_{\boldsymbol{X}}$  in the numerator of eq. \ref{eq:testStat1} is:
\begin{equation}
\label{eq:projMatrix}
\begin{split}
\boldsymbol{P}_{\boldsymbol{X}} &= \boldsymbol{X} \left( \boldsymbol{X}^{\text{T}}\boldsymbol{X} \right)^{-1}\boldsymbol{X}^{\text{T}}, \,\, \text{where:}
\\
\boldsymbol{X} &= \boldsymbol{H}\left( \boldsymbol{H}^{\text{T}}\boldsymbol{H}\right)^{-1}\boldsymbol{A}^{\text{T}}.
\end{split}
\end{equation}
The rank-3 projector matrix  $\boldsymbol{P}_{\boldsymbol{H}}$ $=$ $\boldsymbol{H} \left( \boldsymbol{H}^{\text{T}}\boldsymbol{H} \right)^{-1}\boldsymbol{H}^{\text{T}}$ projects vectors onto the subspace spanned by $\boldsymbol{H}$, $\text{span}\{ \boldsymbol{H} \}$. Projector matrix $\boldsymbol{P}_{\boldsymbol{H}}^{\perp}$ $=$ $\boldsymbol{I}$ - $\boldsymbol{P}_{\boldsymbol{H}}$ in the denominator of eq. \ref{eq:testStat1} then projects vectors onto the space orthogonal to $\text{span}\{ \boldsymbol{H} \}$. We thereby interpret the screening statistic $L_{\boldsymbol{\theta}}\left( \boldsymbol{\mathcal{R}} \right)$ in eq. \ref{eq:testStat1} as a scaled ratio of the radiation pattern energy due to non-VDS faulting (note structure of $\boldsymbol{A}$), divided by the residual radiation pattern energy not attributable to the Rayleigh wave model (the system $\boldsymbol{H}$). Probability theory predicts that this quotient has a noncentral-$F$ distribution with one, and $N-3$ degrees of freedom. The density for $L_{\boldsymbol{\theta}}\left( \boldsymbol{\mathcal{R}} \right)$ is then shaped by a scalar noncentrality parameter $\Lambda$ if $\mathcal{R}_{k}$ includes Gaussian noise ($1 \le k \le N$) and the radiation pattern is non-circular. We write its distributional dependence as $L_{\boldsymbol{\theta}}\left( \boldsymbol{\mathcal{R}} \right)$ $\sim$  $\mathcal{F}_{1,N-3}\left(L,  \Lambda>0 \right)$ and the cumulative, noncentral $\mathcal{F}$ distribution for $L_{\boldsymbol{\theta}}\left( \boldsymbol{\mathcal{R}} \right)$ as:
\begin{equation}
\label{eq:Ldistrib}
\begin{split}
\text{CDF} \left\{ L_{\boldsymbol{\theta}}\left( \boldsymbol{\mathcal{R}} \right) \right\} &= F_{1,N-3}\left(L, \Lambda  \right)
\end{split}
\end{equation}
These distributions have a well-defined variance when their second degree of freedom exceeds four, so that the screening statistic $L_{\boldsymbol{\theta}}\left( \boldsymbol{\mathcal{R}} \right)$ is only well-defined when $N\ge8$. This sampling implies that at least two sensors sample each quadrant of the Rayleigh radiation pattern (given full azimuthal coverage by sensors), which mitigates spatial aliasing of four-lobed radiation patterns. This statistic is identical in form to the subspace detector \citep{Scharf19941} that is often applied in seismic monitoring, and has completely analogous interpretations for waveform detection. Consequently, many standard detection theory texts document the properties and analytical form for $F_{1,N-3}\left(L, \Lambda  \right)$, and computational packages like \texttt{MATLAB}  include routines to calculate random variables and parameters of the noncentral-$F$ distribution. 

Crucially, this theory shows that the scalar $\Lambda$ in eq. \ref{eq:Ldistrib} completely quantifies the screening capability of $L_{\boldsymbol{\theta}}\left( \boldsymbol{\mathcal{R}} \right)$ to test between $\mathcal{H}_{0}$ and $\mathcal{H}_{1}$ at fixed $N$; eq. \ref{eq:genFormNonCentral} shows its general analytical form. Here, we use those results to write $\Lambda$ as the product of one factor that depends on deployment azimuth $\psi$ and a second factor that depends on source and noise parameters:
\begin{equation}
\label{eq:factorLambda}
\begin{split}
\Lambda &= \underbrace{ \cfrac{1}{ \boldsymbol{A} \left[ \boldsymbol{H}^{\text{T}} \boldsymbol{H}\right]^{-1} \boldsymbol{A}^{\text{T}} }}_{\text{Deployment}} \cdot \underbrace{\cfrac{\left( DS + SS \right)^{2}}{\sigma_{M}^{2} + \sigma_{\mathcal{R}}^{2}}}_{\text{faulting SNIR}}.
\end{split}
\end{equation}
This factorization shows that $\Lambda$ is the product of a scalar term $\left( \boldsymbol{A} \left[ \boldsymbol{H}^{\text{T}} \boldsymbol{H}\right]^{-1} \boldsymbol{A}^{\text{T}}\right)^{-1}$ that stores the $N$ sensor deployment locations, and a term $\left( DS + SS \right)^{2}/(\sigma_{M}^{2} + \sigma_{\mathcal{R}}^{2})$ that quantifies the non-VDS faulting signal energy, relative to noise and random process variance. It is analogous to the signal-to-noise plus interference ratio (SNIR) of waveform processing. Explosive or VDS sources that trigger no Rayleigh waves include parameters $\sigma_{\mathcal{R}}^{2}$ $=$ $DS$ $=$ $SS$ $=$ $0$, and $\Lambda$ $=$ $0$ (as expected). Strike-slip faults maximize $\Lambda$ among other faulting sources, given equal sensor placement, because $\left( DS + SS \right)^{2}/\left( \sigma_{M}^{2} +\sigma_{\mathcal{R}}^{2}\right)$ $\le$ $M_{0}^{2}/\left( \sigma_{M}^{2} + \sigma_{\mathcal{R}}^{2}\right)$ (see eq. \ref{eq:tectRelease}). Importantly, eq. \ref{eq:testStat1} and eq. \ref{eq:factorLambda} demonstrate that discriminating between a circular and non-circular Rayleigh wave radiation pattern under the Case I assumptions is entirely equivalent to a signal detection problem. We further note the random process variance $\sigma_{\mathcal{R}}^{2}$ acts to simply inflate the effective noise variance and reduce any observed, non-VDS faulting signal SNR under $\mathcal{H}_{1}$. This asymmetry in variance between the null and alternative hypotheses means that a high SNR faulting signal with high random signal variance is indistinguishable from a lower SNR faulting signal with zero random signal variance. We will assume $\sigma_{\mathcal{R}}^{2} = 0$ hereon, so that $\Lambda$ and $L_{\boldsymbol{\theta}}\left( \boldsymbol{\mathcal{R}} \right)$ are each simplified from their more general forms. This simplification preserves our physical insight into the radiation pattern testing problem without adding cumbersome notation. The faulting SNIR then reduces to just SNR.

Under these simplifying (but still rather general) assumptions, the probability $\text{Pr}_{D}$ of detecting any general faulting signal, and then deciding that $\boldsymbol{\mathcal{R}}$ records a non-circular radiation pattern is the probability $\text{Pr}_{D}$ of measuring a value of $L_{\boldsymbol{\theta}}\left( \boldsymbol{\mathcal{R}} \right)$ that exceeds a threshold $\eta$ (eq. \ref{eq:testStat1}):
\begin{equation}
\label{eq:PrD}
\text{Pr}_{D} = 1 - F_{1,N-3}\left(\eta, \Lambda>0  \right).
\end{equation}
We select a threshold $\eta$ for deciding a radiation pattern is non-circular using the Neyman-Pearson criteria. This criteria uses a constant false-attribution probability $\text{Pr}_{FA}$ to invert for $\eta$:
\begin{equation}
\label{eq:PrFA}
\begin{split}
\text{Pr}_{FA} &=  1 - F_{1,N-3}\left(\eta, \Lambda=0  \right), \,\, (\text{or})
\\
\eta &= F_{1,N-3}^{-1}\left( 1 - \text{Pr}_{FA}, \Lambda=0 \right)
\end{split}
\end{equation}
where we set $\text{Pr}_{FA}$ to an acceptable value (e.g., $10^{-3}$) and $F_{1,N-3}^{-1}\left( \bullet \right)$ is the inverse of the central $F$ distribution with one and $N-3$ degrees of freedom. The screening probability $\text{Pr}_{D}$ in eq. \ref{eq:PrD} monotonically increases with $\Lambda$ at fixed $N$ and $\eta$. The locus of points $\left(\Lambda, \text{Pr}_{D}\right)$ thereby defines performance curves that measure the screening capability of the decision rule in eq. \ref{eq:testStat1} versus SNR of the faulting signal at fixed $\text{Pr}_{FA}$ (Fig. \ref{fig:ncfpdf}). 
Deployment constraints and tectonic settings can each limit the permissible parameter space of $\Lambda$ to a subset $\mathcal{S}\left( \Lambda \right)$. These limits include the azimuthal deployment range of sensor numbers and locations that quantify the leftmost factor in eq. \ref{eq:factorLambda}, and variability in the source focal mechanism, which quantifies the rightmost factor in eq. \ref{eq:factorLambda}. In such cases, we estimate the expected screening capability $\bar{\text{Pr}}_{D} $ of eq. \ref{eq:testStat1} as the average of the detection rate $\text{Pr}_{D}$ over such parametric constraints. We use eq. \ref{eq:PrD} to write this average as:
\begin{equation}
\label{eq:PrDbar}
\begin{split}
\bar{\text{Pr}}_{D}=   \lvert \mathcal{S}\left( \Lambda \right) \rvert^{-1} \displaystyle \sum_{\lvert \mathcal{S}\left( \Lambda \right) \rvert}1 -  F_{1,N-3}\left(\eta, \Lambda  \right).
\end{split}
\end{equation}
in which $\lvert \mathcal{S}\left( \Lambda \right) \rvert$ is the size of the set (e.g., the number of realizations in a Monte Carlo simulation). The sum in eq. \ref{eq:PrDbar} equates to an integration when the subset of permissible values of $\Lambda$ is a continuum (e.g., $\lvert \psi \rvert$ $\le$ $3\pi / 4$) and gives a marginal distribution for $\bar{\text{Pr}}_{D}$ (e.g.,  $\bar{\text{Pr}}_{D}$ $=$ $4/\left(3\pi\right) \int_{-3\pi/4}^{+3\pi/4} d \psi \left( 1 -  F_{1,N-3}\left(\eta, \Lambda  \right) \right)$. 
%
\begin{figure}
 \centering
\includegraphics[width=\textwidth]{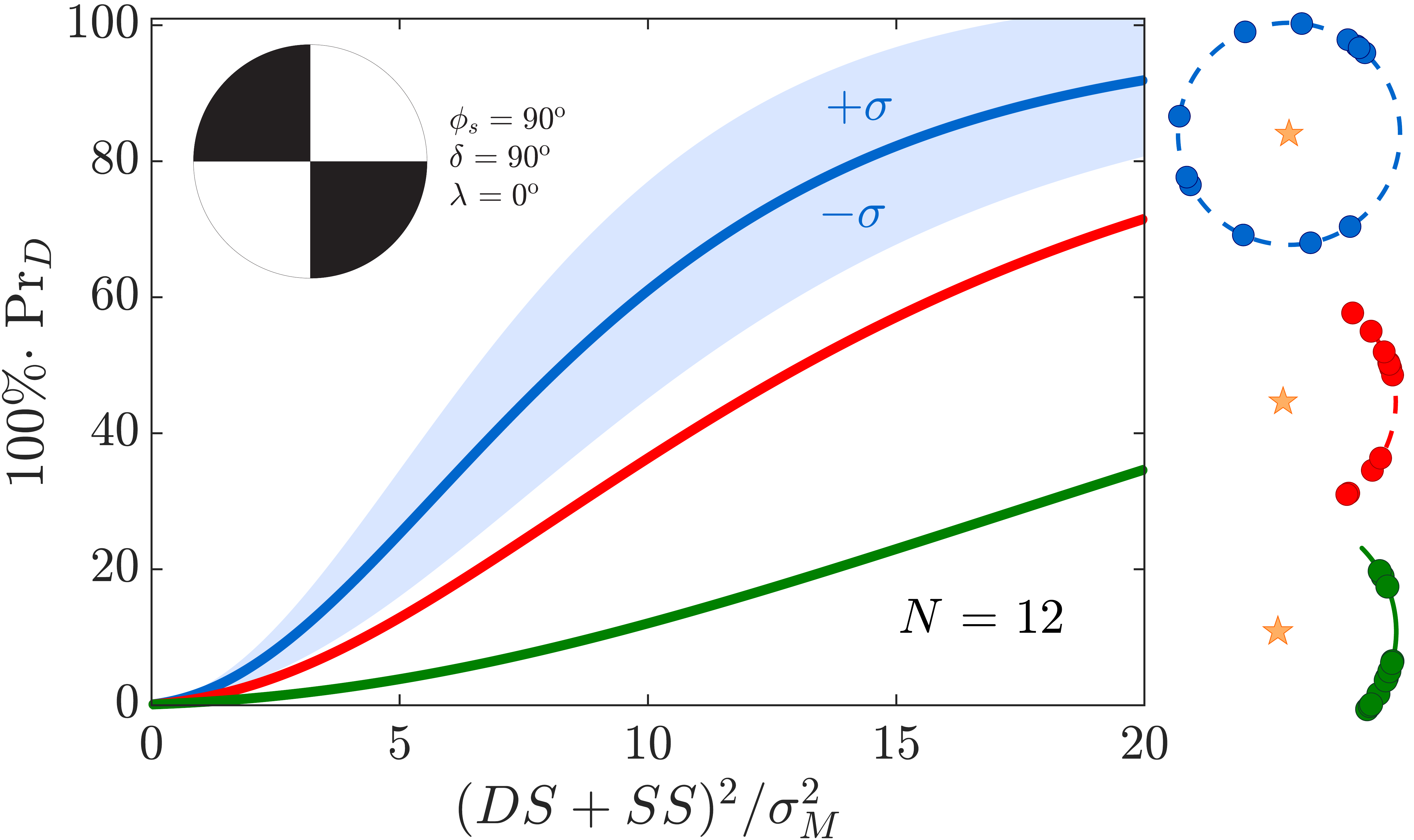}
\captionof{figure}[]
{\narrower Averaged screening curves (eq. \ref{eq:PrDbar}), plotted against faulting signal SNR (horizontal axis;  eq. \ref{eq:factorLambda}). Each curve presents an estimate of $100\%\times$ the probability $\text{Pr}_{D}$ (vertical axis) that the decision rule in eq. \ref{eq:testStat1} will correctly screen a strike-slip earthquake (beach ball, top left) from an explosion. The test statistic $L_{\boldsymbol{\theta}}\left( \boldsymbol{\mathcal{R}} \right)$ (eq. \ref{eq:testStat1}, left-hand side) for each curve associates with $N$ $=$ $12$ sensor azimuths that sample the Rayleigh wavefield $\boldsymbol{\mathcal{R}}$ under distinct deployment constraints (right); all thresholds $\eta$ correspond to $\text{Pr}_{FA}$ $=$ $10^{-3}$ (eq. \ref{eq:PrFA}). The topmost curve (blue) presents an average of $\lvert \mathcal{S}\left( \Lambda \right) \rvert$ $=$ $100$ individual decision rule curves that each measure screening performance when $N$ $=$ $12$ sensors uniformly sample random azimuths along the right, topmost dashed circle that surrounds a source (orange star). The shaded region indicates $\pm$ one standard deviation estimate ($\pm \sigma$) from the mean curve. The second curve (red) presents a similar average of 100 curves that associate with sensors deployed with uniformly random azimuth along the red dashed arc at right surrounding the same source; in this case, the azimuthal range is restricted to $2\pi/3$ radians. The third curve (green) shows a similar average, but for sensor azimuths restricted to $\pi/2$ radians (green dashed arc). The filled circles superimposed on the dashed deployment arcs at right show particular, random sensor locations in each case. Some circular sensor markers overlap.}
 \label{fig:perfCurvesStrikeSlip}
 \end{figure}
 %
 \begin{figure}
 \centering
\includegraphics[width=\textwidth]{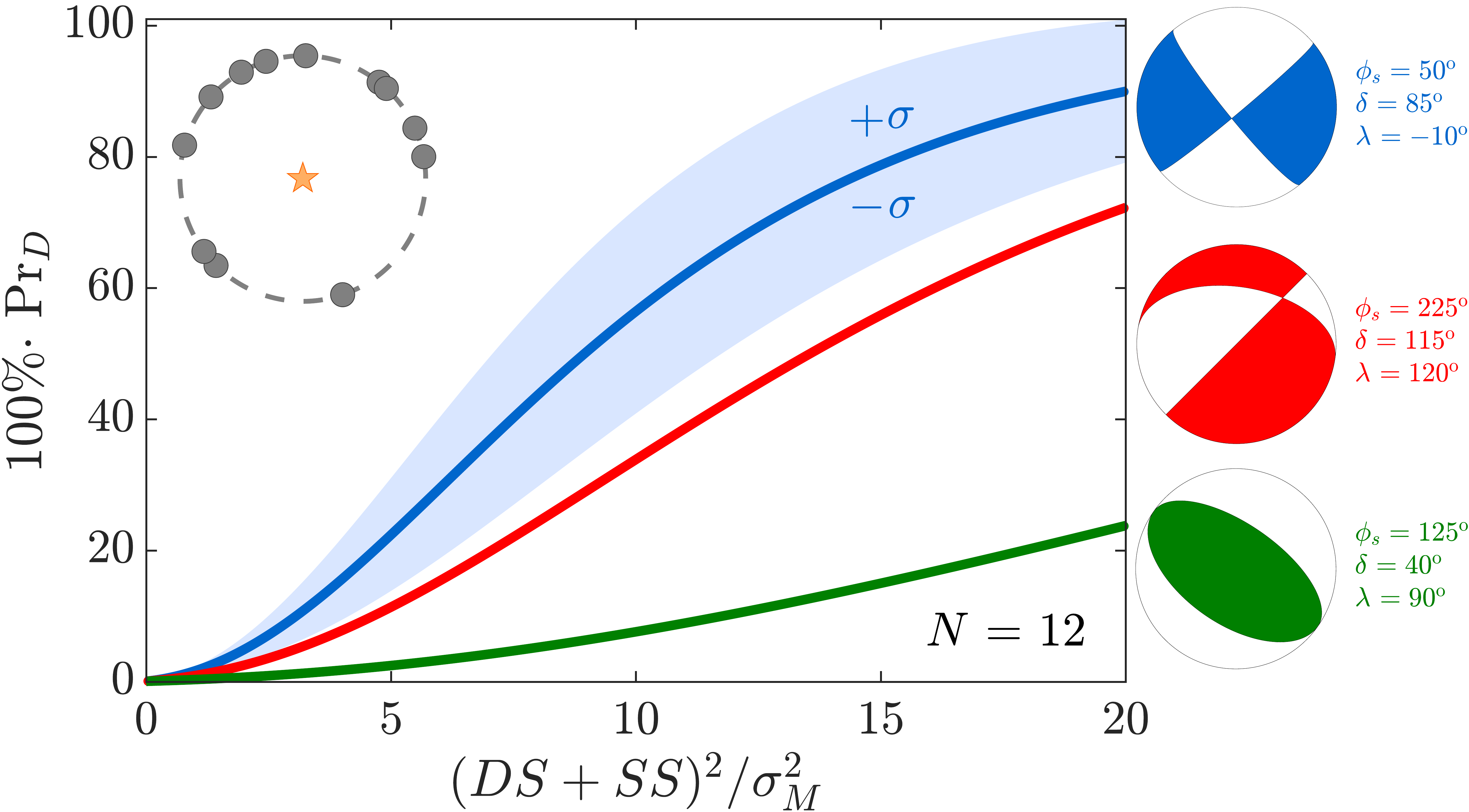}
\captionof{figure}[]
{\narrower Similar to Fig. \ref{fig:perfCurvesStrikeSlip}, but curves now present $100\%\times$ the probability $\text{Pr}_{D}$ (vertical axis) that eq. \ref{eq:testStat1} will screen earthquakes with three distinct focal mechanisms from an explosion, using  $N$ $=$ $12$ sensors, all plotted against faulting signal SNR (horizontal axis;  eq. \ref{eq:factorLambda}). Each curve is an average of 100 screening curves (eq. \ref{eq:PrDbar}) that associate with sensor deployments that uniformly sample all azimuths along the dashed circle centered at the hypocenter of a shallow source (orange star); the filled circles mark a particular, random deployment. The topmost curve (blue) presents such an average for faulting source like that of ``Event 1'' from the Rock Valley earthquake; the second curve (red) presents a similar average for faulting source like that of the Kara Sea event; and the bottom curve presents a similar average for a faulting source like that of the 2003 Lop Nor event. The shaded region indicates $\pm$ one standard deviation from the top curve mean. All curves correspond to $\text{Pr}_{FA}$ $=$ $10^{-3}$.}
 \label{fig:perfCurvesFullCirc}
 \end{figure}
 %
 \begin{figure}
 \centering
\includegraphics[width=\textwidth]{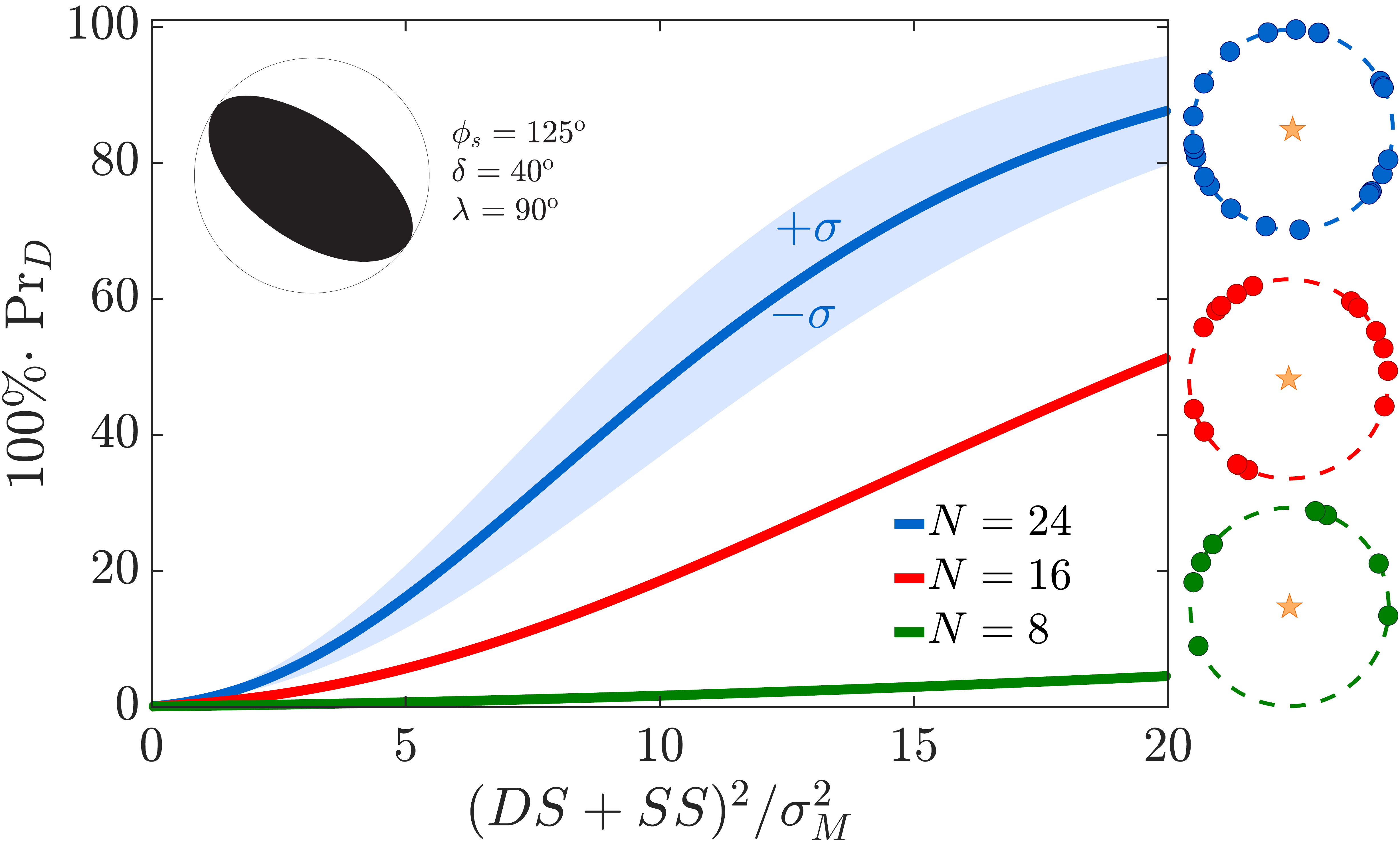}
\captionof{figure}[]
{\narrower Similar to Fig. \ref{fig:perfCurvesStrikeSlip}, but curves now estimate $100\%\times$ the probability that eq. \ref{eq:testStat1} (vertical axis) will screen earthquakes whose focal mechanisms match that of the 2003 Lop Nor event (beachball, upper left) from an explosion, using three distinct sensor deployments without azimuthal constraint. As in Fig. \ref{fig:perfCurvesStrikeSlip} and Fig. \ref{fig:perfCurvesFullCirc}, each curve presents an average of 100 curves (eq. \ref{eq:PrDbar}) that associate with sensors deployed with uniformly random azimuth along the dashed circles (circular markers, right) surrounding the source (orange star). All curves correspond to a $\text{Pr}_{FA}$ $=$ $10^{-3}$ false attribution probability and plot against faulting signal SNR (horizontal axis;  eq. \ref{eq:factorLambda}). Filled sensor circles mark a particular, random deployment.}
 \label{fig:perfCurvesIncrN}
 \end{figure}

Fig. \ref{fig:perfCurvesStrikeSlip}, Fig. \ref{fig:perfCurvesFullCirc} and Fig. \ref{fig:perfCurvesIncrN} each show qualitatively similar, mean screening curves (eq. \ref{eq:PrDbar}) that each index distinct deployment and source parameters. Fig. \ref{fig:perfCurvesStrikeSlip} shows the probability that eq. \ref{eq:PrDbar} will screen strike-slip earthquakes from explosions with $N$ $=$ $12$ sensors deployed over three distinct azimuthal ranges against faulting signal (pictured at right). Each of these three curves show the dependence of screening probability on the faulting signal SNR $\left(DS+SS\right)^{2}/\sigma_{M}^2$, averaged over 100 individual screening curves. Scalar $\Lambda$ uniformly samples receiver deployments confined to these azimuthal ranges where source parameters are $\delta$ $=$ $\frac{\pi}{2}$, $\lambda$ $=$ $0$, $(DS + SS)^{2}$ $=$ $M_{0}$, and faulting SNR $=$ $M_{0}^{2}/\sigma_M^{2}$. The three curves demonstrate that screening probability markedly decreases as azimuthal gap increases with a fixed number of stations, even as station density over deployment azimuth (sensor number per arc length) increases. Shading about the top blue curve associated with full azimuthal sampling marks the one-standard deviation ($\pm \sigma$) variability.

Fig. \ref{fig:perfCurvesFullCirc} illustrates that earthquake signals triggered by other focal mechanisms are more difficult to screen from explosions relative to strike-slip events ($\delta$ $=$ $\pm\pi/2$, $\lambda$ $=$ $0, \pi$). Specifically, eq. \ref{eq:testStat1} illustrates predicted curves that quantify the capability of statistic $L_{\boldsymbol{\theta}}\left( \boldsymbol{\mathcal{R}} \right)$ to screen distinct earthquakes from explosions using the same deployment strategy with $N$ $=$ $12$ sensors. In this case, each curve is an average of $100$ screening curves that randomly sample a circular instrument deployment around three different sources (focal mechanisms at right). These curves demonstrate that the radiation pattern test screens non-strike-slip earthquakes from explosions less effectively than strike-slip earthquakes, even with identical station coverage. Seismic events with a Rayleigh wave radiation pattern like that of the 2003 Lop Nor tectonic event that is sampled by an unrestricted deployment of $N$ $=$ $12$ sensors is particularly difficult to screen from an explosion.  Shading about the curve that associates with full azimuthal sampling marks the $\pm \sigma$ variability in screening sources like Event 1 from the Rock Valley earthquake.

An increase in sensor deployments ($N$ $>$ $12$) correspondingly increases screening probability between non-VDS faulting and explosion sources. Fig. \ref{fig:perfCurvesIncrN} shows three distinct averages of $100$ screening curves that each associate with three distinct sensor deployments ($N$ $=$ $8,16,24$). Each deployment randomly samples a circle centered about a shallow source like that of the 2003 Lop Nor tectonic event. In this case, a screening statistic that uses $N$ $=$ $24$ sensors can screen non-VDS faulting from explosion sources at roughly the same mean rate as a statistic that uses $12$ receivers to screen sources like Event 1 from the Rock Valley earthquake sequence (Fig. \ref{fig:perfCurvesFullCirc}). While a $24$ sensor deployment screens shallow sources like that show in Fig. \ref{fig:perfCurvesIncrN} with relatively low variance, the test statistic obviously requires twice the number of observations. 

\subsection{A Note on Screening Curve Interpretation}
\label{sec:errorInterp}
Our interpretation of screening curves that compare probabilities like $\bar{\text{Pr}}_{D}$ against faulting signal SNIR require comment (see Fig. \ref{fig:perfCurvesStrikeSlip}, Fig. \ref{fig:perfCurvesFullCirc} and Fig. \ref{fig:perfCurvesIncrN}). The binary decision rules necessarily decide between a continuum of source types that are not identically  explosions or faults. For example, the decision rule in eq. \ref{eq:testStat1} will screen data that records Rayleigh motion sourced by an explosion that superimposes with non-zero contributions from a non-VDS faulting signal with a probability that much greater than the false-attribution probability $\text{Pr}_{FA}=10^{-3}$. A proper false attribution probability constraint that considers an acceptable amount of faulting signal with an explosion would more appropriately marginalize out some faulting effects with a prior on $\Lambda$ from this continuum \citep{Berger19871}. We address such source-type errors and ambiguities in Section \ref{sec:errors} of our Discussion. The binary hypothesis testing theory that we use here is standard, however, and we proceed with the caveat that neither the $\mathcal{H}_{0}$ nor $\mathcal{H}_{1}$ in our models fully capture the continuum of source type hypotheses. 

We next explore if more optimal sensor placement can improve screening rates of target sources that are particularly challenging to screen from cylindrically symmetric sources.
\subsection{Case I application: optimal deployment of auxiliary receivers} \label{sec:OptimalDeployment}
We suppose an existing $N\ge8$ receiver network with limited azimuthal coverage requires an auxiliary receiver to maximize its capability to screen shallow non-VDS sources from explosions, through application of $L_{\boldsymbol{\theta}}\left( \boldsymbol{\mathcal{R}} \right)$ as a discriminant. To develop an objective function for this discriminant, we add a row $\left[ 1 , \cos( 2 \psi ) - c ,  \sin( 2 \psi) \right]$ to $\boldsymbol{H}$ and make a new matrix $\boldsymbol{H}_{+1} $. We then maximize $\text{Pr}_{D}$ over additional sensor deployment azimuths $\psi_{+1}$ that we estimate as $\hat{\psi}_{+1}$. Such an optimal auxiliary sensor location estimate supplements the existing network to form an updated $N+1$ network that maximizes the probability of detecting a non-VDS signal in the radiation pattern data $\boldsymbol{\mathcal{R}}$. Specifically, this new observation point updates the discriminant in eq. \ref{eq:testStat1}, while eq. \ref{eq:PrFA} updates the threshold $\eta$ to maintain the same false attribution probability $\text{Pr}_{FA}$. The optimal solution $\hat{\psi}_{+1}$ depends on both $\eta$ and $\Lambda$, and maximizes $\text{Pr}_{D}$ where the total differential $d \text{Pr}_{D}$ peaks:
\begin{equation}
\label{eq:totalDiff}
\begin{split}
\hat{\psi}_{+1} &= \displaystyle \arg\!\max_{ \psi \vert \mathcal{H}_{1}  } \left\{  d \text{Pr}_{D} \right\}  
\\
&=  \displaystyle \arg\!\max_{ \psi \vert \mathcal{H}_{1}  } \left\{ \frac{ \partial \text{Pr}_{D}}{ \partial \eta} \Delta \eta + \frac{ \partial \text{Pr}_{D} }{ \partial \Lambda} \Delta \Lambda \right\}.
\end{split}
\end{equation}
The decision threshold decreases by an increment $\Delta \eta$:
\begin{equation}
\begin{split}
\Delta \eta = &F_{1,N-2}^{-1}\left( 1 - \text{Pr}_{FA}, \Lambda=0 \right) - \\
&F_{1,N-3}^{-1}\left( 1 - \text{Pr}_{FA}, \Lambda=0 \right),
\end{split}
\end{equation}
whereas the faulting signal changes by increment  $\Delta \Lambda$:
\begin{equation}
\label{eq:LambdaIncr}
\begin{split}
\Delta \Lambda = &\cfrac{ \left(DS + SS\right)^{2} } { \sigma_{M}^{2} } \times
\\
&\left[ \cfrac{1} { \boldsymbol{A} \left[ \boldsymbol{H}_{+1}^{\text{T}} \boldsymbol{H}_{+1}\right]^{-1} \boldsymbol{A}^{\text{T}} }  
-
\cfrac{1} { \boldsymbol{A} \left[ \boldsymbol{H}^{\text{T}} \boldsymbol{H}\right]^{-1} \boldsymbol{A}^{\text{T}} }  \right].
\end{split}
\end{equation} 
The $\arg\!\max$ argument on the right-hand-side eq. \ref{eq:totalDiff} is a directional derivative. It is maximal at parameter increments $\left[ \Delta \eta,\, \Delta \Lambda \right]$ in the direction of gradient $\nabla \text{Pr}_{D}$ $=$ $\left[\frac{\partial \text{Pr}_{D}} {\partial \eta},\, \frac{\partial \text{Pr}_{D}} {\partial \Lambda }\right]$, where $\text{sign} \left\{ \frac{\partial \text{Pr}_{D}} {\partial \eta} \right\}$ $=$ $-1$ and $\text{sign} \left\{ \frac{\partial \text{Pr}_{D}} {\partial \Lambda} \right\}$ $=$ $+1$, and only increment $\Delta \Lambda$ depends on $\psi_{+1}$. Moreover, the quadratic forms in \ref{eq:LambdaIncr} are both strictly positive, since $\left[ \boldsymbol{H}^{\text{T}} \boldsymbol{H}\right]^{-1} $ is positive definite. This collective system structure implies that the bracketed difference between the geometric factors is always non-negative (e.g., adding more sensors cannot reduce the source screening probability), so that:
\begin{equation}
\begin{split}
\label{eq:varHatNplusOne}
\hat{\psi}_{+1} &=  
\displaystyle \arg\!\max_{ \psi \vert \mathcal{H}_{1}  } \left\{ d \text{Pr}_{D} \right\} 
\\
&= \displaystyle \arg\!\max_{ \psi \vert \mathcal{H}_{1}  } \left\{ \Delta \Lambda \right\}
\\
&= 
\displaystyle \arg\!\min_{ \psi \vert \mathcal{H}_{1}  }  \left\{  \boldsymbol{A} \left[ \boldsymbol{H}_{+1}^{\text{T}} \boldsymbol{H}_{+1}\right]^{-1} \boldsymbol{A}^{\text{T} } \right\}.
\end{split}
\end{equation}
The last term of eq. \ref{eq:varHatNplusOne} defines the objective function for Rayleigh wave sampling. It omits explicit dependence on data $\boldsymbol{\mathcal{R}}$. Therefore, the optimal strike-relative deployment azimuth estimate $\hat{\psi}_{+1}$ for a receiver when $\sigma_{M}^{2}$ is unknown depends only on current receiver placement and the relative seismic wave structure ($\alpha$ and $\beta$). Other values of sensor placement where $\boldsymbol{A} \left[ \boldsymbol{H}_{+1}^{\text{T}} \boldsymbol{H}_{+1}\right]^{-1} \boldsymbol{A}^{\text{T}}$ $<$ $1$ increase $\Lambda$ from its prior value (eq. \ref{eq:factorLambda}). Auxiliary sensor azimuths where $\boldsymbol{A} \left[ \boldsymbol{H}_{+1}^{\text{T}} \boldsymbol{H}_{+1}\right]^{-1} \boldsymbol{A}^{\text{T}}$ $>$ $1$ decrease $\Lambda$ from its prior value. To quantify the change in screening power, we reapply eq. \ref{eq:PrFA} to estimate the updated threshold $\eta$ and reapply eq. \ref{eq:PrD} to compute source screening rates with $N+1$ sensors; we then update the decision rule (eq. \ref{eq:testStat1}).
%
\begin{figure*}
 \centering
\includegraphics[width=1\textwidth]{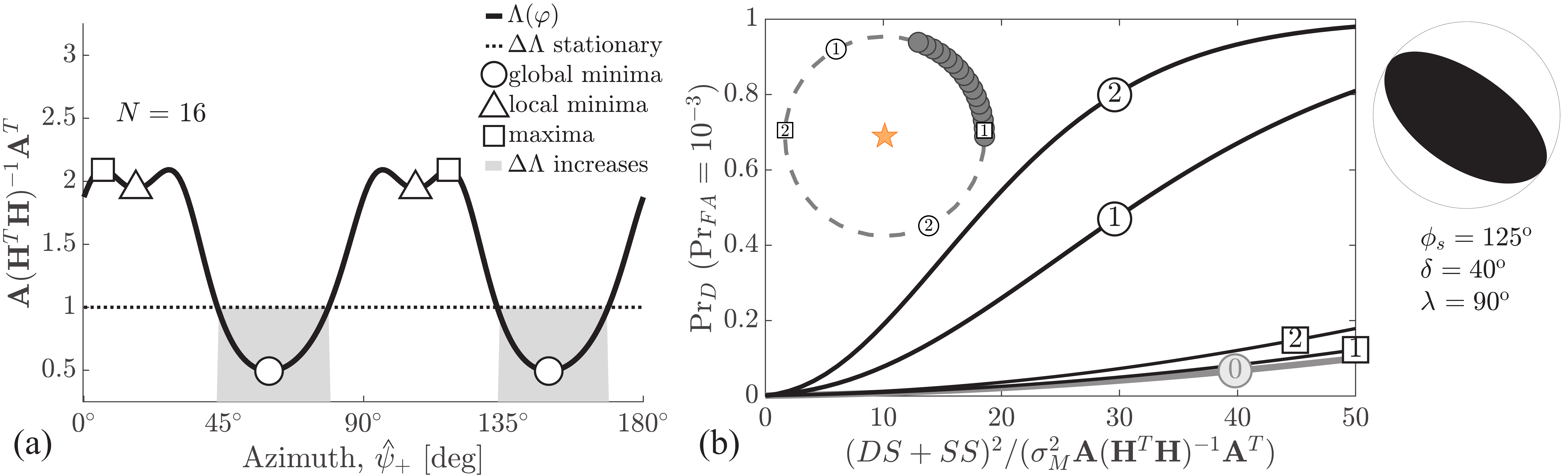}
\captionof{figure}[]
{\narrower Optimal and sub-optimal azimuthal solutions for extra sensor locations to supplement an existing sensor deployment. (\textbf{a}): Solution curves to eq. \ref{eq:varHatNplusOne} that show the dependence of optimization parameter $\Lambda$ on the geometric deployment factor (left factor in eq. \ref{eq:factorLambda}; vertical axis) on azimuth $\psi$ (horizontal axis). Points below the dashed, ``stationary'' line marks azimuthal regions where both $\Delta \Lambda$ and $\Delta \eta$ increase. The global minima (circles) mark solutions that maximize source screening parameter $\Lambda$. The local minima (triangles) mark solutions for the best location of an extra receiver that is near the worst deployment location (squares). (\textbf{b}): Screening probability curves $\text{Pr}_{D}$ (vertical axis) for the decision rule in eq. \ref{eq:varHatNplusOne} when deployments include one and two extra sensors to supplement an existing $N$ $=$ $16$ station network (gray markers, upper left), compared against the scaled faulting signal (horizontal axis). The radiation pattern source has the same focal mechanism as the 2003 Lop Nor tectonic event (beach ball at top right). The thick gray curve labeled with a gray, circled $0$ marks the performance of the screening statistic that associates with the original instrument deployment (gray circles). The other numbered curves show the screening performance of the optimally deployed $N$ $=$ $17$ and $N$ $=$ $18$ networks. These curves associate with the sensor locations that we depict with deployment schemes (upper left) and minima in (\textbf{a}). Squares mark curves that show worse-case deployment azimuths that correspond to maxima on the curve in (\textbf{a}).}
 \label{fig:optimSolCurves}
 \end{figure*}

Fig. \ref{fig:optimSolCurves}a shows $\pi$-periodic optimal and sub-optimal solutions for extra receiver locations that supplement $16$ existing receivers uniformly deployed along a $75^{\circ}$ azimuthal arc (gray circles, Fig. \ref{fig:optimSolCurves}b). The objective function shows local and global extrema of $ \boldsymbol{A} \left[ \boldsymbol{H}_{+1}^{\text{T}} \boldsymbol{H}_{+1}\right]^{-1} \boldsymbol{A}^{\text{T}}$. The maxima occur at deployment azimuths that minimize $\Lambda$ and therefore the probability of screening binary source types; these sensor locations provide the poorest improvement in source-type screening. Local minima (triangles) provide a locally optimal sensor placements near these worst-case sensor locations. These local minima solutions may apply to scenarios in which deployment outside restricted azimuthal ranges is impossible (e.g., inaccessible areas). Our global solutions (circles) confirm expectation that optimal receiver deployment should include points that include no current receiver coverage. These optimal deployment locations are not orthogonal to the mid-point of the azimuthal arc, but depend on the relative values of the body wave speeds (parameter $c$). Fig. \ref{fig:optimSolCurves}b shows screening curves for the decision rule \ref{eq:testStat1} when a receiver network contains the backbone of existing sensors (inset gray circles), supplemented with poorly placed sensors (squared one and two), versus optimally placed extra sensors (circled 1 and 2). The optimally placed sensor that supplements this $16$ sensor backbone network forms a $17$ sensor network the improves screening rates of the faulting source by $> 6\times$ that of the backbone network, for sources with the greatest faulting signal shown (circled 1). An additional, optimally placed sensor (circled 2) provides a greater gain in screening performance, relative to an $18$ sensor network that includes two poorly placed auxiliary sensors (squared 2). The screening performance our test achieves with this azimuthally limited, 18 sensor arrangement outperforms average test that uses the $24$ sensor deployment in Fig. \ref{fig:perfCurvesIncrN}.

Lastly, we emphasize that the update to $\eta$ maintains a consistent false-attribution probability for the discriminant before and after adding any additional receivers. Failure to update this threshold when $\Delta \Lambda$ $<$ $1$ could lead an observer to erroneously conclude that adding receivers at particular values of $\psi_{+1}$ (shaded domain of Fig. \ref{fig:optimSolCurves}(a)) can reduce the screening power of the same test.
\subsection{Case I application: threshold faulting moments and missed detections}
We now quantify the largest faulting source signal (as defined by $\left( DS + SS \right)^{2}/\sigma_{M}^{2}$) that could be mistakenly attributed to an explosion, at a fixed false attribution probability $\text{Pr}_{FA}$. This application is particular important to underground explosion monitoring applications because it provides fundamental physical limits on the the discrimination power of eq. \ref{eq:HypotTest-vec} tests to screen explosive from non-VDS faulting sources. 

\begin{figure}
 \centering
\includegraphics[width=\textwidth]{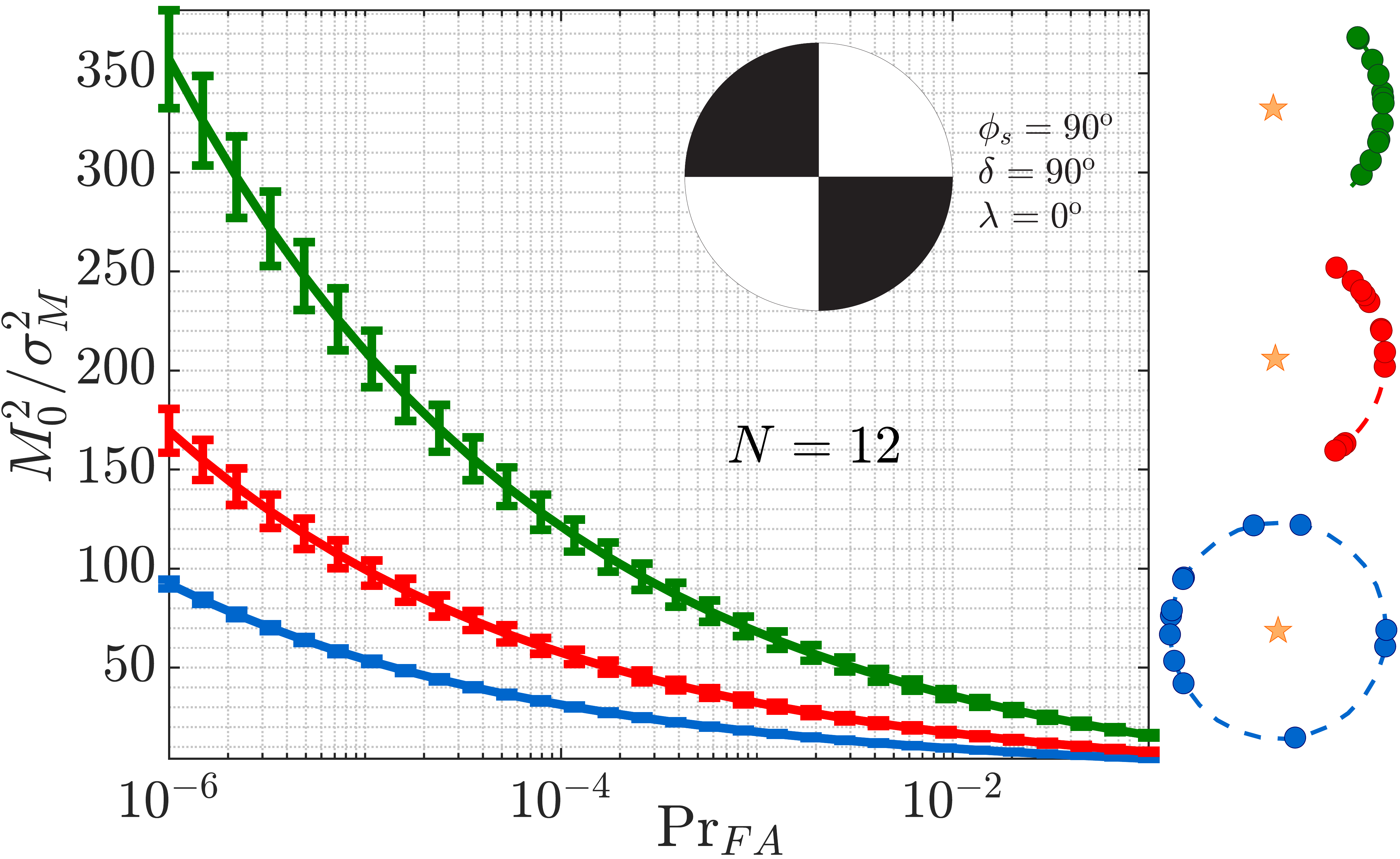}
\captionof{figure}[]
{\narrower Three solutions curves to eq. \ref{eq:PrMLambdaSrc} that quantify the largest strike-slip faulting source signal that the decision rule in eq. \ref{eq:testStat1} can mis-identify to originate from an explosion with probability $1-\text{Pr}_{D}$  $=$ $0.1$. In each case, $N=12$ sensors confined to three distinct azimuthal deployment ranges (at right) sample the strike-slip Rayleigh wave radiation pattern (beach ball inset). Each curve compares faulting signal SNR (vertical axis) against false attribution probability (horizontal axis), and error bars measure standard error from 100 random sensor deployments. Filled circles (right) mark a particular, random deployment (we use standard error here because standard deviation markers reduce readability).}
 \label{fig:PrM}
 \end{figure}

To compute this signal, we first note that the probability of mistaking a non-VDS faulting source for an explosive source is equivalent to the event that $L_{\boldsymbol{\theta}}\left( \boldsymbol{\mathcal{R}} \right)$ $<$ $\eta$ when $\mathcal{H}_{1}$ is true (eq. \ref{eq:testStat1}). Here, threshold $\eta$ relates to the CDF under hypothesis $\mathcal{H}_{0}$  through eq. \ref{eq:PrFA}. The probability of miss-attributing a non-VDS faulting source to an explosion then equates to $1-\text{Pr}_{D}$ and relates to the faulting signal through a nonlinear equation that requires inversion of the noncentral-$F$ CDF for $\Lambda$:
\begin{equation}
\label{eq:PrMiss}
1-\text{Pr}_{D} = F_{1, N-3} \left(  \eta; \Lambda  = \frac{1}{\boldsymbol{A} \left[ \boldsymbol{H}^{\text{T}} \boldsymbol{H}\right]^{-1} \boldsymbol{A}^{\text{T}} } \frac{\left( DS + SS \right)^{2}}{\sigma_{M}^{2}} \right)
\end{equation}
The term $1-\text{Pr}_{D}$ is equivalent to a ``miss" in waveform detection theory. The faulting signal SNR that solves eq. \ref{eq:PrMiss} relates to noncentrality parameter estimate $\hat{\Lambda}$ through:
\begin{equation}
\begin{split}
\label{eq:PrMLambda}
\hat{\Lambda} &=  
\displaystyle \arg\!\min_{ \Lambda } \left\{ \lvert 1-\text{Pr}_{D} -  F_{1, N-3} \left(  \eta; \Lambda  \right) \rvert \right\}.
\end{split}
\end{equation}
Our estimate for the source term of $\hat{\Lambda}$, as measured by a given sensor deployment locations $\psi_{m}$ ($m=1,2,\cdots,N$) that are stored in $\boldsymbol{H}$, is:
\begin{equation}
\begin{split}
\label{eq:PrMLambdaSrc}
\widehat{ \frac{\left( DS + SS \right)^{2}}{\sigma_{M}^{2}} } =  \left( \boldsymbol{A} \left[ \boldsymbol{H}^{\text{T}} \boldsymbol{H}\right]^{-1} \boldsymbol{A}^{\text{T}} \right) \hat{\Lambda}.
\end{split}
\end{equation}
Fig. \ref{fig:PrM} illustrates three solution curves to eq. \ref{eq:PrMLambdaSrc} for a strike-slip faulting signal when $1-\text{Pr}_{D}$ $=$ $0.1$ (for a $0.9$ correct attribution probability), which depicts averages over $100$ random, azimuthally constrained sensor deployments. These solutions provide a lower bound on threshold source size, since $\Lambda$ is largest for strike-slip events ($\left( DS + SS \right)^{2} / \sigma_{M}^{2}$ $=$ $M_{0}^{2}/\sigma_{M}^{2}$). Vertical differences between distinct curves suggest that false attribution rates of a non-VDS faulting signal to an explosion source (via eq. \ref{eq:testStat1}) increases with azimuthal gap. 
%

\begin{figure*}
 \centering
\includegraphics[width=1\textwidth]{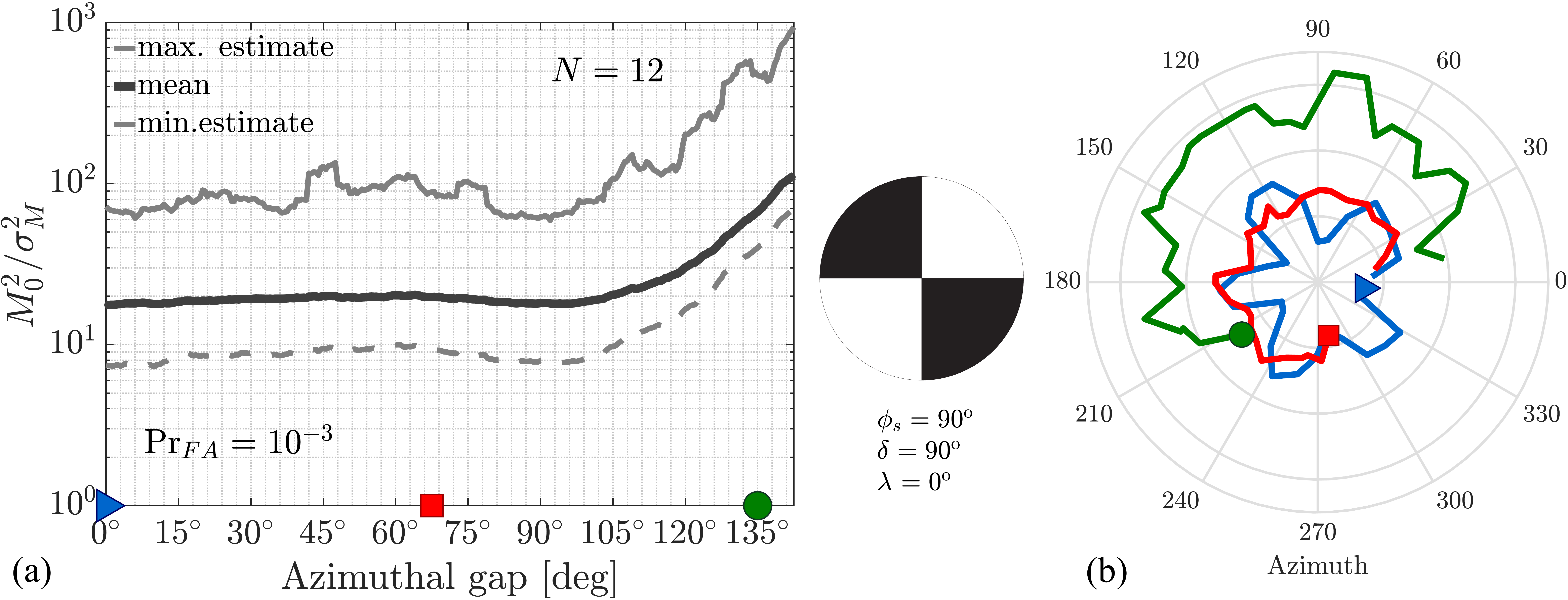}
\captionof{figure}[]
{\narrower Solutions to eq. \ref{eq:PrMLambdaSrc} for strike-slip sources (Case I). \textbf{(a)}: Moving window statistics of threshold faulting SNR values (vertical axis) that associate with $1-\text{Pr}_{D}$ $=$ $0.1$ and $N=12$ sensors and that are confined by a grid of azimuthal gaps (horizontal axis). The noise free strike-slip beach ball appears at right. Markers on the horizontal axis associate with Rayleigh wave radiation patterns in (b). \textbf{(b)}: Three noisy radiation patterns of a strike-slip source (eq. \ref{eq:HypotTest-vec}) with three corresponding azimuthal gaps in sensor sampling. The decision rule in eq. \ref{eq:testStat1} samples these radiation patterns at $N$ $=$ $12$ locations to falsely attribute the source of each pattern to an explosion, with probability $1-\text{Pr}_{D}$ $=$ $0.1$. A four-lobe Rayleigh wave radiation pattern is evident in the case that sensors sample the radiation field without azimuthal constraint (blue pattern). We emphasize that the decision rule in eq. \ref{eq:testStat1} mistakes this pattern as circular (with a $0.1$ probability).} 
 \label{fig:PrMrad}
 \end{figure*}

Fig. \ref{fig:PrMrad}(a) better quantifies this gap effect on screening explosive from strike-slip sources. Each family of curves show the maximum, mean and minimum solutions to eq. \ref{eq:PrMLambdaSrc} per azimuthal gap value. The vertical range of SNR values between each curve measures the variability that results from the $100$ randomly placed sensors. In particular, these solution curves indicate that strike-slip source size (SNR $=$ $M_{0}^{2}/\sigma_{M}^{2}$) increases exponentially as deployment gap increases beyond $90^{\circ}$ (equivalent to a $270^{\circ}$ azimuthal coverage). Moreover, the variability in such threshold estimates is asymmetric: the log-scale maximum threshold estimates (top curve) depart from the mean (middle curve) significantly more than do log-scale minimum threshold estimates (bottom curve). Noisy radiation patterns that associate with these solutions in Fig. \ref{fig:PrMrad}(b) show that the decision defined by eq. \ref{eq:testStat1} can mistake (with probability  $1-\text{Pr}_{D}$ $=$ $0.1$) the subtle four-lobed shape for a noisy circular radiation pattern, even with full azimuthal sensor coverage. 
%
\subsection{Case II: $\sigma_{R}$ and $\boldsymbol{\theta}$ unknown; $\sigma_{M}$ known}
We progress from Case I and now assume that a sensor network records a Rayleigh wave radiation pattern with unknown random process variability ($\sigma_{\mathcal{R}}^{2}$ in eq. \ref{eq:HypotTest-vec}), unknown faulting parameters ($\boldsymbol{\theta}$ in eq. \ref{eq:systemMatrix}), and a stochastic variance $\sigma_{M}^{2}$ that is either known, or has a significantly lower estimator variance than than that of $\hat{\sigma}_{\mathcal{R}}^{2}$. This case applies to idealized monitoring scenarios in which an underground testing complex colocates with a shallow fault of unknown geometry and radiation variability, but with well-characterized stochastic variability $\sigma_{M}^{2}$ that is perhaps quantified from historical records of proximal explosions. Sensor networks that record a low magnitude seismic event that originates from this test site faces the same challenge as that in Case I, but the screening statistic has a distinct form. We treat Case II in less detail than Case I because it appears less practically applicable than Case I while remaining more complicated. Appendix \ref{sec:CaseIIpdf} derives several results that we state here. For Case II, the log-GLRT equivalent to eq. \ref{eq:maxLikeGLR} is:
\begin{equation}
\begin{split}
L_{\boldsymbol{\theta}}^{(II)}\left( \boldsymbol{\mathcal{R}} \right)
\,
&\underset{\mathcal{H}_{0}}
{ \overset{\mathcal{H}_{1}}
{\gtrless}}
\,
\eta \quad \text{where:}
\\ 
L_{\boldsymbol{\theta}}^{(II)}\left( \boldsymbol{\mathcal{R}} \right) &= \ln \left[ \cfrac{ \displaystyle \max_{ {\boldsymbol{\theta}, \, \sigma_{\mathcal{R}}^{2}}  } \{ \, f_{\boldsymbol{\mathcal{R}}} \left(\boldsymbol{\mathcal{R}}; \mathcal{H}_{1} \right) \, \} }{\,  \displaystyle \max_{ {\boldsymbol{\theta}}  } \{ f_{\boldsymbol{\mathcal{R}}} \left( \boldsymbol{\mathcal{R}}; \mathcal{H}_{0} \right) \, \} } \right],
\end{split}
\label{eq:maxLikeGLR_2}
\end{equation}
where eq. \ref{eq:nullPdf}  and eq. \ref{eq:nullPdf0} define $f_{\boldsymbol{\mathcal{R}}} \left( \boldsymbol{\mathcal{R}}; \mathcal{H}_{0} \right)$ and eq. \ref{eq:altPdf} defines $f_{\boldsymbol{\mathcal{R}}} \left( \boldsymbol{\mathcal{R}}; \mathcal{H}_{1} \right)$. The MLE for $\sigma_{\mathcal{R}}^{2}$  is (algebra omitted):
\begin{equation}
\begin{split}
\hat{\sigma}_{R}^{2} &= \cfrac{\vert \vert  \boldsymbol{\mathcal{R}} - \boldsymbol{H} \hat{\boldsymbol{\theta}}_{1}  \vert \vert^{2} }{N} - \sigma_{M}^{2}
\\
&= \cfrac{\vert \vert  \boldsymbol{P}_{\boldsymbol{H}}^{\perp} \boldsymbol{\mathcal{R}} \vert \vert^{2} } {N} - \sigma_{M}^{2}
\end{split}
\label{eq:MLEsigmaRsquared}
\end{equation}
in which the text that follows eq. \ref{eq:projMatrix} defines the projector matrix $\boldsymbol{P}_{\boldsymbol{H}}^{\perp}$. Our eq. \ref{eq:defineTheta} defines the MLEs $\hat{\boldsymbol{\theta}}_{k}$ for parameter vectors $\boldsymbol{\theta}_{k}$ ($k=0,1$). We input these parameters to eq. \ref{eq:MLEsigmaRsquared} and reuse the symbol $L_{\boldsymbol{\theta}}\left( \boldsymbol{\mathcal{R}} \right)$ to write the resultant, equivalent GLRT statistic (algebra omitted):
\begin{equation}
\label{eq:subpsaceStat}
\begin{split}
L_{\boldsymbol{\theta}}\left( \boldsymbol{\mathcal{R}} \right) &=   \cfrac{\vert \vert  \boldsymbol{\mathcal{R}} - \boldsymbol{H} \hat{\boldsymbol{\theta}}_{0}  \vert \vert^{2} }{\sigma_{M}^2} - N \ln \left[   \cfrac{\vert \vert  \boldsymbol{P}_{\boldsymbol{H}}^{\perp}  \boldsymbol{\mathcal{R}}  \vert \vert^{2} }{\sigma_{M}^2}  \right].
\end{split}
\end{equation}
We use the MLEs for $\hat{\boldsymbol{\theta}}_{k}$ to express $L_{\boldsymbol{\theta}}\left( \boldsymbol{\mathcal{R}} \right)$ as a sum of two statistically independent random variables ($X$ and $Y$):
\begin{equation}
\label{eq:CaseIITestStat}
L_{\boldsymbol{\theta}}\left( \boldsymbol{\mathcal{R}} \right) =   \underbrace{\cfrac{ \bigr \Vert \boldsymbol{P}_{\boldsymbol{H}}^{\perp} \boldsymbol{\mathcal{R}} \bigr \Vert^{2} }
{\sigma_{M}^2} - N \ln \left[  \cfrac{ \bigr \Vert \boldsymbol{P}_{\boldsymbol{H}}^{\perp} \boldsymbol{\mathcal{R}} \bigr \Vert^{2} }{\sigma_{M}^2}  \right]}_{X}
+
\underbrace{\cfrac{ \bigr \Vert \boldsymbol{P}_{\boldsymbol{X}} \boldsymbol{\mathcal{R}} \bigr \Vert^{2} } { \sigma_{M}^2  }}_{Y}.
\end{equation}
We interpret the first quadratic form ($\Vert \boldsymbol{P}_{\boldsymbol{H}}^{\perp} \boldsymbol{\mathcal{R}} \Vert^{2} / / \sigma_{M}^2 $) in eq. \ref{eq:CaseIITestStat} as the residual radiation pattern energy not attributable to the Rayleigh wave model, relative to random noise variance. The second term ($\Vert \boldsymbol{P}_{\boldsymbol{X}} \boldsymbol{\mathcal{R}} \Vert^{2}  / \sigma_{M}^2 $) is energy due to non-VDS faulting, relative to the same noise variance. The decision rule that tests the function of these two terms in eq. \ref{eq:CaseIITestStat} is:
\begin{equation}
\begin{split}
\label{eq:testStat2}
\cfrac{ \bigr \Vert \boldsymbol{P}_{\boldsymbol{H}}^{\perp} \boldsymbol{\mathcal{R}} \bigr \Vert^{2} }
{\sigma_{M}^2} - N \ln \left[  \cfrac{ \bigr \Vert \boldsymbol{P}_{\boldsymbol{H}}^{\perp} \boldsymbol{\mathcal{R}} \bigr \Vert^{2} }{\sigma_{M}^2}  \right]
+
\cfrac{ \bigr \Vert \boldsymbol{P}_{\boldsymbol{X}} \boldsymbol{\mathcal{R}} \bigr \Vert^{2} } { \sigma_{M}^2  }
\underset{\mathcal{H}_{0}}{ \overset{\mathcal{H}_{1}}
{\gtrless}}
\,
\hat{\eta}.
\end{split}
\end{equation}
%
\begin{figure}
 \centering
\includegraphics[width=\textwidth]{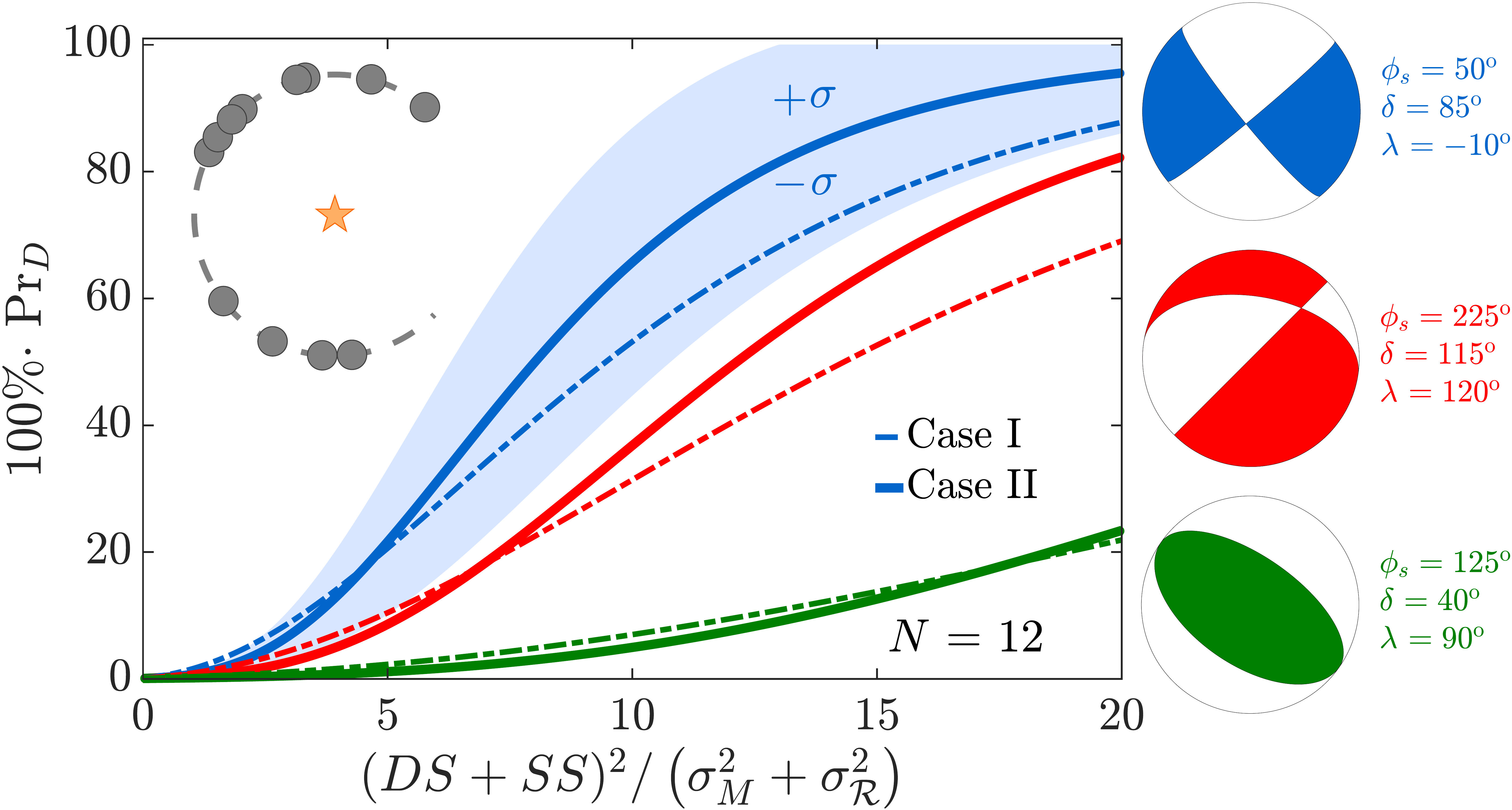}
\captionof{figure}[]
{\narrower Screening curves for Case II (thick, solid curves) compared with Case I test statistics (thin, dashed curves). The plot format is similar to Fig. \ref{fig:perfCurvesFullCirc}: Case I and Case II curves present $100\%\times$ the probability $\text{Pr}_{D}$ (vertical axis) that eq. \ref{eq:testStat1} and  will screen earthquakes with three distinct focal mechanisms from an explosion, using  $N$ $=$ $12$ sensors, all plotted against faulting signal SNIR (horizontal axis;  eq. \ref{eq:factorLambda}). Each curve is an average of 100 screening curves (eq. \ref{eq:PrDbar}) that associate with sensor deployments that uniformly sample azimuths with a $90^{\circ}$ gap along the dashed circle centered at the hypocenter of the shallow source (orange star); the filled circles mark a particular, random deployment. The topmost curve (blue) presents such an average Case II screening curve for an earthquake with a focal mechanism that matches that of ``Event 1'' from the Rock Valley earthquake; the second curve (red) presents a similar average for an earthquake with a focal mechanism like that of the Kara Sea event; and the bottom curve presents a similar average for an earthquake with a focal mechanism like that of the 2003 Lop Nor event. The shaded region indicates $\pm$ one standard deviation ($\pm \sigma$) from the top curve mean. All curves correspond to $\text{Pr}_{FA}$ $=$ $10^{-3}$.}
 \label{fig:CaseIIvCaseI}
\end{figure}

The threshold $\hat{\eta}$ in eq. \ref{eq:testStat2} is strictly an estimate that maintains a false-attribution probability $\text{Pr}_{FA}$, consistent with Case I. We describe our estimation strategy for this threshold in Appendix \ref{sec:CaseIIpdf} (see eq. \ref{eq:interpCaseIIThr}). To compute the screening performance for Case II (SNIR versus $\text{Pr}_D$), we note that the terms $X$ and $Y$ are independent, piecewise invertible functions. In particular, the term $X$ is piecewise invertible in argument $s$ $=$ $\Vert \boldsymbol{P}_{\boldsymbol{H}}^{\perp} \boldsymbol{\mathcal{R}} \Vert^{2} /\sigma_{M}^2$ since function $x(s)$ $=$ $s$ $-$ $N \ln(s)$ monotonically decreases over $s$ $<$ $N$ and monotonically increases when $s$ $>$ $N$. The ratio $s$ $\sim$ $\chi_{N-3}^{2}(0)$, where $\chi_{N-3}^{2}(0)$ is chi-squared distribution function with $N-3$ degrees of freedom and a zero noncentrality parameter; the distributional form of $s$ is identical for both circular and non-circular radiation patterns. Unlike the Case I test statistic, these PDFs are well defined if $N\ge4$ sensors rather than eight sensors. The ratio $Y$ $\sim$ $\chi_{1}^{2}(\Lambda)$ is a chi-squared distribution function with $1$ degrees of freedom and a nonzero noncentrality ($\Lambda$ $>$ $0$) under $\mathcal{H}_{1}$. We use these results to compute the PDF of the sum $Z$ $=$ $X$ $+$ $Y$ through a variable transformation on the original, statistically independent random variables. Given that $X$ and $Y$ respectively have PDFs $f_{X}\left( x\right)$ and $f_{ Y }\left( y\right)$, the PDF $f_{Z}\left( z; \mathcal{H}_{i} \right)$ for $Z$ is their convolution $f_{X} \left( x\right) \ast f_{ Y }\left( y\right)$ under hypothesis $\mathcal{H}_{i}$ ($i=0,1$). Omitting the explicit algebra, we symbolize the PDF $f_{Z}\left( z; \mathcal{H}_{i} \right)$ as:
\begin{equation}
\label{eq:Zstat}
f_{Z}\left( z; \mathcal{H}_{i} \right) = \int_{-\infty}^{\infty}  \underbrace{f_{\chi_{N-3}^{2}}\left( s(x) \right) \biggr \vert \cfrac{ds}{dx} \biggr \vert}_{f_{X}\left(x\right)} \cdot \underbrace{f_{ \chi_{1}^{2} }\left(z- x \right)}_{f_{ Y }\left(z-x\right)} dx
\end{equation}
We compute eq. \ref{eq:Zstat} in the frequency domain with the characteristic function method (eq. \ref{eq:fZPdf}). The PDF $f_{ \chi_{1}^{2}}(\bullet)$ includes a finite noncentrality parameter under the alternative hypothesis ($i=1$). The theory we detail in Appendix \ref{sec:CaseIIpdf} demonstrates that this parameter equates to the same scalar $\Lambda$ in eq. \ref{eq:factorLambda} that we derived from Case I. This means that $\Lambda$ completely quantifies the screening capability of $L_{\boldsymbol{\theta}}\left( \boldsymbol{\mathcal{R}} \right)$ to test between $\mathcal{H}_{0}$ and $\mathcal{H}_{1}$. The PDF $f_{X}\left(x\right)$ is identical under $\mathcal{H}_{0}$ and $\mathcal{H}_{1}$. We interpret its effect on $f_{Z}\left( z; \mathcal{H}_{i} \right)$ as a smoothing window in the convolution operation in eq. \ref{eq:Zstat} that is independent of the Rayleigh wave radiation pattern model. 

Fig. \ref{fig:CaseIIvCaseI} shows the screening power of the Case II test statistic compared against that of the Case I test statistic, for three source types and a fixed deployment constraint (an azimuthal gap of $90^{\circ}$) for $N=12$ sensors. The graph format is identical to that shown by Fig. \ref{fig:perfCurvesStrikeSlip}, Fig. \ref{fig:perfCurvesFullCirc} and Fig. \ref{fig:perfCurvesIncrN}, in which each curve represents an average (eq. \ref{eq:PrDbar}) over 100 deployment realizations. Each such realization effectively recomputes the deployment factor $\left( \boldsymbol{A} \left[ \boldsymbol{H}^{\text{T}} \boldsymbol{H}\right]^{-1} \boldsymbol{A}^{\text{T}} \right)^{-1}$ of $\Lambda$ and re-convolves the PDFs $f_{ Y }\left( \bullet \right)$ and $f_{ X }\left( \bullet \right)$ in the frequency domain as a spectral product. We document the additional numerical details that we implemented to construct the screening curves shown by Fig. \ref{fig:CaseIIvCaseI} in text following eq. \ref{eq:interpCaseIIThr}. 

Importantly, these curves demonstrate that the Case II decision rule provides a superior source-screening capability, at least for the same faulting signal SNIR and thresholds required to maintain a false attribution probability of $\text{Pr}_{FA}$ $=$ $10^{-3}$. We expect the Case II test statistic to provide such an improved, average performance because the Case I statistic additionally requires an MLE of the noise variance ($\hat{\sigma}_{M}^{2}$) under $\mathcal{H}_{0}$; in general, test statistics with unknown parameters show poorer screening performance than competing test statistics with fewer unknown parameters \cite[pg. 195]{Kay19981}. The variability in screening performance with sensor placement over permissible azimuths (note the $90^{\circ}$ gap), however, is comparable between both Case I and Case II (note shading). Both tests are also similar in the relative order of screening rates with focal mechanism. In particular, the Case II statistic shows a qualitatively similar ordering in screening probability when compared to the Case I statistic. The Case II decision rule (eq. \ref{eq:testStat2}) that tests radiation pattern data that resembles the Rock Valley earthquake (for example) shows a much higher probability of correctly screening such events from isotropic sources than, say, data that records an event that resembles the 2003-03-13 Lop Nor earthquake. 
\section{Discussion}
Our binary hypothesis test (eq. \ref{eq:HypotTest-vec}) quantifies an observer's idealized ability to screen shallowly buried explosion sources from non-VDS faulting sources. These tests effectively measure deviations from circularity in Rayleigh wave radiation patterns that any faulting terms induce at the source location (Fig. \ref{fig:HypTest}). While the binary test provides only a commensurate, binary decision on the significance of any non-circular contributions to a measured radiation pattern, it affords significant insight into an observer's capability to distinguish source types. 

\subsection{Case I}
We first consider our results under the general assumptions of Case I, when $\sigma_{R}^{2}$ $=$ $0$:
\begin{itemize}
\item
An observer can use azimuthally distributed receivers that sample a noisy Rayleigh wave radiation pattern to form a test statistic and decision rule (eq. \ref{eq:testStat1}). This rule screens explosive from non-VDS faulting sources at a given threshold $\eta$ that maintains a fixed false attribution probability $\text{Pr}_{FA}$.
\\
\item
We interpret this screening statistic as a ratio between radiation pattern energy due to non-VDS faulting and residual radiation pattern energy that is not captured by our Rayleigh wave model. This test statistic has a noncentral $F$-distribution that quantifies this observer's screening capability, and is well-defined only when an observer deploys eight or more receivers.
\\
\item
Our test between circular versus non-circular Rayleigh radiation patterns is equivalent to a general Gaussian signal detection problem (eq. \ref{eq:testStat1}). The product of a faulting SNR term and a scalar deployment term (eq. \ref{eq:factorLambda}) defines the test screening power. The first term measures the squared sum of the dip- and strike- slip faulting components, relative to moment noise: $(DS + SS)^{2}/\sigma_{M}^{2}$. The deployment term $\left( \boldsymbol{A} \left[ \boldsymbol{H}^{\text{T}} \boldsymbol{H}\right]^{-1} \boldsymbol{A}^{\text{T}}\right)^{-1}$ depends on azimuthal sensor distribution and the body wave speeds of the host medium (parameter $c$).
\\
\item 
Strike-slip sources provide the highest faulting signal strength $M_{0}^{2}/\sigma_{M}^{2}$ for a given faulting moment $M_{0}$. Eq. \ref{eq:testStat1} therefore screens strike-slip from explosive and vertical dip-slip faulting sources with the highest probability $\text{Pr}_{D}$ $=$ $1 - F_{1,N-3}\left(\eta, M_{0}^{2}/\sigma_{M}^{2}  \right)$, among other sources with the same faulting moment $M_{0}$ and constant false attribution threshold $\eta$ $=$ $F_{1,N-3}^{-1}\left( 1 - \text{Pr}_{FA}, 0 \right)$. 
 \\
\item Radiation pattern screening probabilities depend on three parameters: azimuthal gap, source radiation pattern, and the number of deployed sensors. Fig. \ref{fig:perfCurvesStrikeSlip}, Fig. \ref{fig:perfCurvesFullCirc} and Fig. \ref{fig:perfCurvesIncrN} show that certain combinations of these three parameter sets produce qualitatively similar radiation pattern screening curves. 
\\ 
\item An observer can use the screening test to estimate the azimuthal placement of additional sensors that supplement an existing network, and which maximize the probability of screening circular from non-circular radiation patterns. Moreover, deploying additional sensors at any azimuth can never reduce screening rates.
\\
\item Lastly, observers have a non-negligible probability of mis-attributing non-circular radiation patterns of faulting sources to explosions when they measure these patterns with sparse sensor deployments. The SNR of such a faulting signal increases exponentially as the azimuthally gap in sensor deployment exceeds $\sim90^{\circ}$. This SNR can exceed an order of magnitude as gaps increase from $0^{\circ}$ to $135^{\circ}$ (Fig. \ref{fig:PrMrad}).
\end{itemize}
We next consider our (more limited) results under the general assumptions of Case II.

\subsection{Case II}
The assumptions under Case II explicitly hypothesize that an unknown, random process superimposes with a deterministic radiation pattern signal and imprints on the total pattern. It also assumes that stochastic variability within the radiation pattern (noise) is effectively known under each hypothesis. This latter assumption remains less practical because the seismic noise field is often insufficiently stationary to be well-characterized, at least at multiple azimuths. We can, however, report several outcomes:
\begin{itemize}
\item
The Case II statistic increases with Rayleigh wave energy sourced by non-VDS faulting, relative to noise power ($\Vert \boldsymbol{P}_{\boldsymbol{X}} \boldsymbol{\mathcal{R}} \Vert^{2}  / \sigma_{M}^2 $). It is non-monotonic in residual energy that is not captured by our Rayleigh wave model and normalized by noise power ($\Vert \boldsymbol{P}_{\boldsymbol{H}}^{\perp} \boldsymbol{\mathcal{R}} \Vert^{2} / \sigma_{M}^2 $).
\\
\item
The screening power of the Case II decision rule (eq. \ref{eq:testStat2}) depends on a scalar function of faulting signal and sensor deployment. This scalar term $\Lambda$ is identical to that under Case I (eq. \ref{eq:factorLambda}).
\\
\item
A spectral product of two PDFs quantifies the screening performance of the Case II test statistic. The resulting PDF does not appear to have a known standard form. It's CDF predicts that the Case II decision rule (eq. \ref{eq:testStat2}) is more effective at screening circular from non-circular radiation patterns than the Case I decision rule (eq. \ref{eq:testStat1}), because it requires fewer MLEs.
\\
\item
The Case II test statistic is (as opposed to Case I) well defined with $N$ $\ge$ $4$ sensors (rather than eight sensors).
\end{itemize}
Both the Case I and Case II results have implications to screen SPE-triggered explosion-plus-damage signals from historical faulting signals that are each sourced in the NNSS testing complex. We suggest that an observer can use $N$ $=$ $12$ sensors deployed around the source with an azimuthal gap of $90^{\circ}$ or less to screen faulting signals that match such historically recorded mechanisms, provided these data have sufficient SNR (Fig. \ref{fig:CaseIIvCaseI}). A confirmation of screening power from such a validation exercise remains necessary if this screening statistic holds efficacy in non-idealized settings.

\subsection{Caveats}
Our results remain subject to at least four more specific caveats. First, our work applied a planar model applicable to local distances that concern records from sources like the SPE or New Hampshire shot series. More established observational and theoretical work that quantifies the imprint of faulting motion on Rayleigh wave radiation patterns (e.g., tectonic release) considers far-regional to teleseismic data. Such Rayleigh waveforms are dominated by long period spectral content and propagation distances of many hundreds to thousands of km. We assert hypothesis tests with our model remain applicable to such global problems. Specifically, we argue that we can rotate the data at each sensor location through multiplication by a unitary matrix to a locally planar coordinate system; this idempotent matrix multiplication does not change the distributional form of such quadratic forms, nor the expected performance of each test statistic.

Second, we did not explore the relative contributions of the explosion versus the CLVD source to the circular part of the radiation pattern in our model. This CLVD term can exceed the explosion term in size, which differs in sign so that Rayleigh waveforms reverse polarity. While the Case I and Case II test statistics exclude any such circular terms ($\Lambda$ excludes $\Delta \bar{R}$), we concede that a zero-mean damage signal may appear in the data as a random process. Physically, this means dilatation that is symmetric about the vertical axis may manifest as a late time CLVD-source. Our linear model does not accommodate such moment-dependent changes in the total data variance. The possible presence of such a CLVD signal motivates our future consideration of the Case II test statistic.

Third, we assumed that noise recorded at distinct sensor azimuths is uncorrelated, largely for mathematical convenience. Correlated noise reduces the effective degree of freedom that parameterize the distribution of each test statistic. This means that our theoretical screening rates will exceed our observed detection rate if our method does not accommodate for noise correlation. We can make such accommodations by computing empirically-validated estimates for the effective degrees of freedom in observed data \cite[eq. A1, eq. A3]{Carr20201}. This argument applies to both Case I and Case II. 

Fourth and lastly, our noisy Rayleigh wave models impart ``ragged edges'' to the graphical radiation pattern representations (Fig. \ref{fig:PrMrad}). These distortions from their deterministic patterns that are sourced by noise may be relatively small when compared to distortions sourced by deterministic focusing or other scattering effects not included in this model (in a real Earth). Such deterministic distortion would shape the radiation pattern from isotropic source (a pure explosion) so that this pattern would appear to have a partial ``lobe''. Our decision rules (eq. \ref{eq:testStat1} or eq. \ref{eq:testStat2}) would then show increased Type 1 errors, that is, a higher probability of falsely deciding that the data include a non-VDS faulting signal (choosing $\mathcal{H}_{1}$ when $\mathcal{H}_{0}$ is true).  Alternatively, scattering of Rayleigh wave energy from surface topography that is comparable to Rayleigh wavelengths might homogenize the radiation pattern from a source with a considerable faulting signal. We expect this homogenization far from the source through energy conservation and divergence arguments: more energy is available to be directed from anti-nodal azimuths to nodal azimuths from any non-zero scattering effects. This example would elevate Type 2 error rates, that is, the probability that our decision rules incorrectly decide that a tectonic source is explosive in origin (choosing $\mathcal{H}_{0}$ when $\mathcal{H}_{1}$ is true); this latter example would imply that eq. \ref{eq:testStat1} will mistake sources with larger faulting signals than shown in Fig. \ref{fig:PrMrad} as explosions more often than we predict. In the language of signal detection, these collective physical processes present in a real Earth will inflate false alarm and missed detection rates over our predicted rates.
%
\begin{figure}
 \centering
\includegraphics[width=\textwidth]{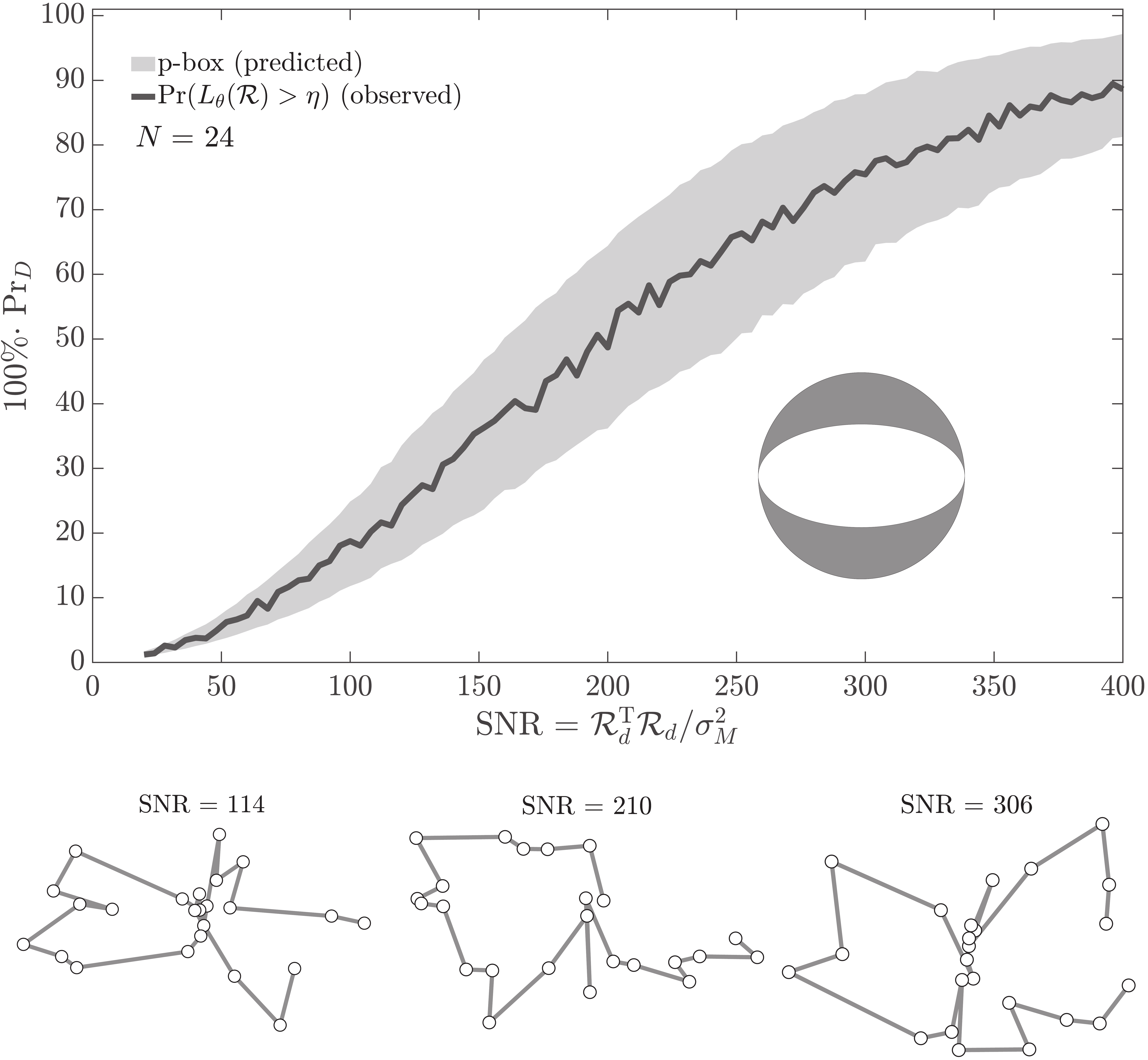}
\captionof{figure}[]
{\narrower A range of predicted screening curves (p-box) superimposed with the mean observed screening curve for the Case I decision rule (eq. \ref{eq:testStat1}), which we apply to a synthetic radiation pattern collected from azimuthal records of Rayleigh waves triggered by an opening surface crack (beachball at center, right; see Section \ref{sec:errors}). Curves present $100\%\times$ the probability $\text{Pr}_{D}$ (vertical axis) that eq. \ref{eq:testStat1} will use  $N$ $=$ $24$ sensors to decide that a fracture source produces a non-circular radiation pattern and includes a faulting signal, plotted against radiation pattern SNR (horizontal axis). The shaded region shows a measure of the range of theoretical screening curves. The curve that marks the lower limit of this range is parameterized by noncentrality parameter $\Lambda$ $=$ $\Vert \boldsymbol{P}_{\boldsymbol{X}} \boldsymbol{\mathcal{R}}_{d} \Vert^{2} /\sigma_{M}^2$ $-$ $\sigma$, where $\sigma$ is the standard deviation marking expected variability of estimates for $\Lambda$, which originate from $\sigma_{M}^2$. The curve that marks the upper limit of this range is similarly parameterized by noncentrality parameter $\Lambda$ $=$ $\Vert \boldsymbol{P}_{\boldsymbol{X}} \boldsymbol{\mathcal{R}}_{d} \Vert^{2} /\sigma_{M}^2$ $+$ $\sigma$. The dark curve presents an average of $2\times10^{4}$ synthetic decision rule counts, that is, the number of times that eq \ref{eq:testStat1} chooses hypothesis $\mathcal{H}_{1}$. All curves correspond to $\text{Pr}_{FA}$ $=$ $10^{-3}$. The synthetic ``ragged edged'' radiation patterns with variance $\sigma_{M}^2$ at bottom associate with the labeled SNR values. White circles mark azimuthal sensor locations.}
 \label{fig:crackRadPatt}
\end{figure}
\subsection{A Physical Source of a Screening Error}
\label{sec:errors}
We explicitly treat an analytical example of an erroneous screening test that we apply to a radiation pattern that is induced by a vertical crack (see comments in Section \ref{sec:errorInterp}) opening within a stratified half-space. This example produces a noncircular radiation pattern without the faulting signal that is modeled by hypothesis $\mathcal{H}_{1}$. To construct this pattern, we first consider the moment density tensor for a shallow displacement discontinuity  \citep[eq. 10.6.6.]{Pujol20031}. We then write the (full) moment tensor for such a discontinuity that is source by a vertically oriented surface crack of area $A$, located at the source point $\boldsymbol{\xi}_{0}$, that opens in tension a distance $\llbracket u(\boldsymbol{\xi}_{0}) \rrbracket$ in the Cartesian $y$ direction (algebra omitted):
\begin{equation}
\label{eq:crackMomentTensor}
\boldsymbol{M} = \rho A \llbracket u(\boldsymbol{\xi}_{0}) \rrbracket \beta^{2}
\left( \begin{array} {ccc}
\frac{\alpha^{2}}{\beta^{2}} - 2  & 0 & 0 \\
0 & \frac{\alpha^{2}}{\beta^{2}}& 0 \\
0 & 0 & \frac{\alpha^{2}}{\beta^{2}} - 2\\
\end{array}\right).
\end{equation}
Eq. \ref{eq:crackMomentTensor} omits the frequency-domain representation for the source time function that would describes the history of the crack-face displacement and appear as a factor of $\boldsymbol{M}$. We note that this moment tensor is a sum of a tensor from an explosion source, and a tensor from a force couple that is aligned with the opening crack. We use the $\boldsymbol{M}$ in eq. \ref{eq:crackMomentTensor} and the radiation pattern that appears in both eq. \ref{eq:displ} and eq. \ref{eq:cartRadPatt} to write the Rayleigh surface displacement $\boldsymbol{u}$ as (again, omitting the source-time function):
\begin{equation}
\label{eq:displaceExpanded}
\begin{split}
\boldsymbol{u}(\boldsymbol{\xi}, \omega) &= \underbrace{\rho A \llbracket u(\boldsymbol{\xi}_{0}) \rrbracket \beta^{2}}_{\text{units: energy}} \left( \frac{3 \alpha^{2} - 4 \beta^{2}}{\alpha^{2}} \right) \underbrace{\left( 1 - \cfrac{\alpha^{2} \cos{2\varphi}}{ 3\alpha^{2}  - 4 \beta^{2} } \right)}_{\text{crack radiation pattern}} \boldsymbol{g}(\boldsymbol{\xi}, \omega).
\end{split}
\end{equation}
The angle $\varphi$ in Eq. \ref{eq:displaceExpanded} measures azimuth from the linear crack face. We pair each term with the deterministic radiation pattern coefficients that associate with trigonometric basis functions in the noisy description of $\mathcal{R}$ in eq. \ref{eq:distrHypotTest}:
\begin{equation}
\label{eq:crackRadPatts}
\begin{split}
\bar{\mathcal{R}} &= \rho A \llbracket u(\boldsymbol{\xi}_{0}) \rrbracket \beta^{2} \left( \frac{3 \alpha^{2} - 4 \beta^{2}}{\alpha^{2}} \right) 
\\
\mathcal{R}_{1} &= \rho A \llbracket u(\boldsymbol{\xi}_{0}) \rrbracket \beta^{2} \cdot \cos \left( 2 \varphi \right)
\\
\mathcal{R}_{2} &= 0 \cdot \sin \left( 2 \varphi \right)
\end{split}
\end{equation}
which have units of moment ($\text{F} \cdot \text{m}$). The term $\rho A \llbracket u(\boldsymbol{\xi}_{0}) \rrbracket \beta^{2} $ represents the elastic energy released by volumetric crack growth. Such sources are common in seismogenic hydrofracture events within glaciers \citep{Carr20201, Taylor20191} and ice sheets \citep{Carmichael20151}.

We now suppose $N$ sensors distributed at distinct azimuths along a circle that is centered at the source crack samples the radiation pattern in eq. \ref{eq:crackRadPatts}, which we store in vector $\boldsymbol{\mathcal{R}}_{d}$ (subscript $d$ indicates deterministic). These sensor azimuths populate the system matrix $\boldsymbol{H}$ in eq. \ref{eq:projMatrix}, and the noncentrality parameter equivalent to eq. \ref{eq:genFormNonCentral} becomes $\Vert \boldsymbol{P}_{\boldsymbol{X}} \boldsymbol{\mathcal{R}}_{d} \Vert^{2} /\sigma_{M}^2$; here, eq. \ref{eq:projMatrix} defines $\boldsymbol{P}_{\boldsymbol{X}}$ and $\sigma_{M}^2$ is Gaussian noise variance. We then add noise with this variance to the fracture-generated radiation pattern and process these synthetic data (now  $\boldsymbol{\mathcal{R}}$ $=$ $\boldsymbol{\mathcal{R}}_{d}$ $+$ $\boldsymbol{n}$, where $\boldsymbol{n}$ is noise) with our Case I test statistic $L_{\boldsymbol{\theta}}\left( \boldsymbol{\mathcal{R}} \right)$ and decision rule (eq. \ref{eq:testStat1}) over a grid of crack release energies and SNR values.  In particular, we considered ratios of $20$ $\le$ $\vert \vert \boldsymbol{\mathcal{R}}_{d} \vert  \vert^{2} / \sigma_{M}^{2}$ $\le$ $400$ and $N=24$ sensors. Fig. \ref{fig:crackRadPatt} shows that the Case I test erroneously screens opening cracks as explosions for low SNR, or as non-VDS faulting sources at higher SNR values. Either decision may be interpreted as an error.


\section{Conclusions and Future Work}
The goal of this work was to determine if a discriminant for shallow source types that tests Rayleigh wave radiation pattern shapes is justified on theoretical grounds. We conclude that this approach is likely not practical in most passive monitoring scenarios that require sampling the Rayleigh wave field over a non-horizontally stratified medium with heterogeneities and substantial topographic relief (a real Earth) for two reasons. First, an observer maintains a good discrimination capability between circular and non-circular radiation patterns only with a large number of sensors with good azimuthal coverage and high SNR in a cylindrically symmetric medium. Second, any distortion to the Rayleigh radiation pattern sourced by deterministic processes (not just noise) will increase Type 1and Type 2 error rates in the decision rules that provide a source-type discrimination capability, over that of the ideal cases.

In particular, our hypothesis tests on radiation pattern shape show our tests achieve good, but idealized screening power over only certain regions of parameter space that include deployment azimuth, focal mechanism, and sensor number. We demonstrated that a Rayleigh wave radiation pattern test that exploits a sufficient number of sensors ($>$ $12$) with limited azimuthal gap ($<90^{\circ}$) to record radiation from sources with moderate strike-slip faulting signals (SNR $> 20$), like the historical Rock Valley Event 1, show a high probability $\text{Pr}_{D}$ of success ($\text{Pr}_{D}$ $>$ $0.9$). The probability of such success diminishes with more ambiguous focal mechanisms that resemble certain historical, shallow earthquakes that locate near the Lop Nor nuclear test site in China. Our further technical developments require validating the screening capability of this method against SPE shots that source near historical, tectonic sources of non-circular Rayleigh wave radiation. To overcome challenges of limited explosion data (single shots), we will perform semi-empirical tests on existing explosion records. These tests will infuse amplitude-scaled Rayleigh waveforms triggered by a known source into long noise records, thousands of times, and will re-compute screening curves from the resultant, semi-synthetic data. This process thereby will provide a set of ``observed'' screening curves that we can directly compare to our predicted screening curves that test radiation pattern circularity. 

In achieving the goal of this work, our hypothesis testing approach did provide two unexpected, ancillary benefits to experimental planning of seismic deployments. First, we can quantify how a particular geographical placement of additional sensors to supplement an existing  dense network can drastically increase our capability to screen the non-VDS faults from the explosive sources.  We conclude from this case that optimal sensor deployment does not depend on data records, only prior sensor locations and the relative compressional versus shear wave speed. Second, our screening statistic quantifies the largest faulting signal that an observer could expect to record and misattribute to an explosion source (for some probability). The size of this source increases rapidly with azimuthal gap in receiver deployments that exceed $90^{\circ}$. Our estimates are applicable to test-ban treaty verification scenarios that require attributing discriminant evidence to source identity, but (as stated) we must first assess if such a test even has marginal screening power with real data.

We concede that there are at least three technical considerations outside the scope of this current study (additional to caveats). First, we did not supplement our radiation pattern tests with parallel analysis of Love wave radiation pattern circularity (present under hypothesis $\mathcal{H}_{1}$). Second, we do not consider far-regional and teleseismic observations that require spherical propagation models. Third, we did not explore how damage processes that contribute to the CLVD source, which can reverse Rayleigh waveform polarity, contribute to the apparent stochastic noise in our data. We note that some evidence suggests that random processes can model vertical axis, cylindrically symmetric damage signals in Rayleigh waveform records. Such processes will contribute to our Case II radiation pattern models.

While this theoretical analysis did not use not waveform data, our results do suggest that radiation pattern shapes provide screening power as a physically-interpretable, source-type discriminant in some idealized cases. Such cases include controlled, laboratory experiments with engineered structures, or natural, horizontally stratified structures that are leveraged in seismic studies elsewhere \citep{Toksoz19711, Biagi19901, Lott20201, Lu20071, Hudson20201, Knox20161}. If future application of our tests with locally recorded data prove successful, we will begin testing it on regional, low frequency data. If our method instead fails, then we will quantify the geophysical source of this disagreement to gain physical insight into how Rayleigh radiation pattern shape can or cannot be tested as a discriminant.
\section{Data and Resources}
Our theoretical study used no data. Fig. \ref{fig:allRadPatts} provides references that document estimates for hypocentral solutions and magnitude estimates for reference sources. \texttt{MATLAB} version 2019b output all computational results and plots. The corresponding author may provide scripts upon request, under the auspices of the United States Department of Energy (US DOE).
\section{Acknowledgments}
This research was funded by the National Nuclear Security Administration, Defense Nuclear Nonproliferation Research and Development (NNSA DNN R\&D). The author acknowledges important interdisciplinary collaboration with scientists and engineers from LANL, LLNL, MSTS, PNNL, and SNL. We thank Dale Anderson, Pat Brug, Leslie Casey, and Brian Paeth for their support. We thank Carene Larmat, Howard Patton, Mike Begnaud and Neill Symons for content suggestions. Don Montoya constructed Fig. \ref{fig:HypTest} at Los Alamos National Laboratory.  We thank one anonymous reviewer for pointing out an energy divergence argument that homogenizes the radiation pattern through scattering in real Earth. We also thank Dr. Mark Fisk for a particularly thorough review and constructive criticism. 
This manuscript has been authored with number LA-UR-20-27299 by Triad National Security under Contract with the U.S. Department of Energy, Office of Defense Nuclear Nonproliferation Research and Development. The United States Government retains and the publisher, by accepting the article for publication, acknowledges that the United States Government retains a non-exclusive, paid-up, irrevocable, world-wide license to publish or reproduce the published form of this manuscript, or allow others to do so, for United States Government purposes.
\bibliographystyle{apalike}
\bibliography{Carmichael-FY2021-GJI-20-0929}
\clearpage
\begin{appendices}
\setcounter{equation}{0}
\renewcommand{\theequation}{\Alph{section}.\arabic{equation}} 
\setcounter{section}{0}
\renewcommand*{\thesection}{\Alph{section}}
\setcounter{figure}{0}
\renewcommand*{\thefigure}{\Alph{section}.\arabic{figure}} 

\section{Deterministic Radiation Pattern Moments} \label{sec:radPatMoments}
We compute deterministic moments of a Rayleigh wave radiation pattern $\mathcal{R}$ that is produced from buried, shallow sources; we emphasize that ``moments'' in this appendix often refer to probabilistic moments, so we explicitly label seismic moments. The probabilistic moments include the (1) deterministic mean $\bar{\mathcal{R}}$ and (2) the deterministic variance $\langle \sigma_{\mathcal{R}}^{2} \rangle$ of $\mathcal{R}$. We also bound their behavior with their root-mean-square (RMS) values $\sqrt{ \langle \mathcal{R}^{2} \rangle }$, and $\langle \sigma_{\mathcal{R}}^{2} \rangle$ to measure the expected deviation of the radiation pattern from $\bar{\mathcal{R}}$ $+$ $\langle \sigma_{\mathcal{R}}^{2} \rangle $. We assume sensors are deployed with equal probability around the source so that $\psi$ is uniformly distributed over $[0,2\pi]$. 
%
\begin{figure*}
\centering
\includegraphics[width=1.0\textwidth]{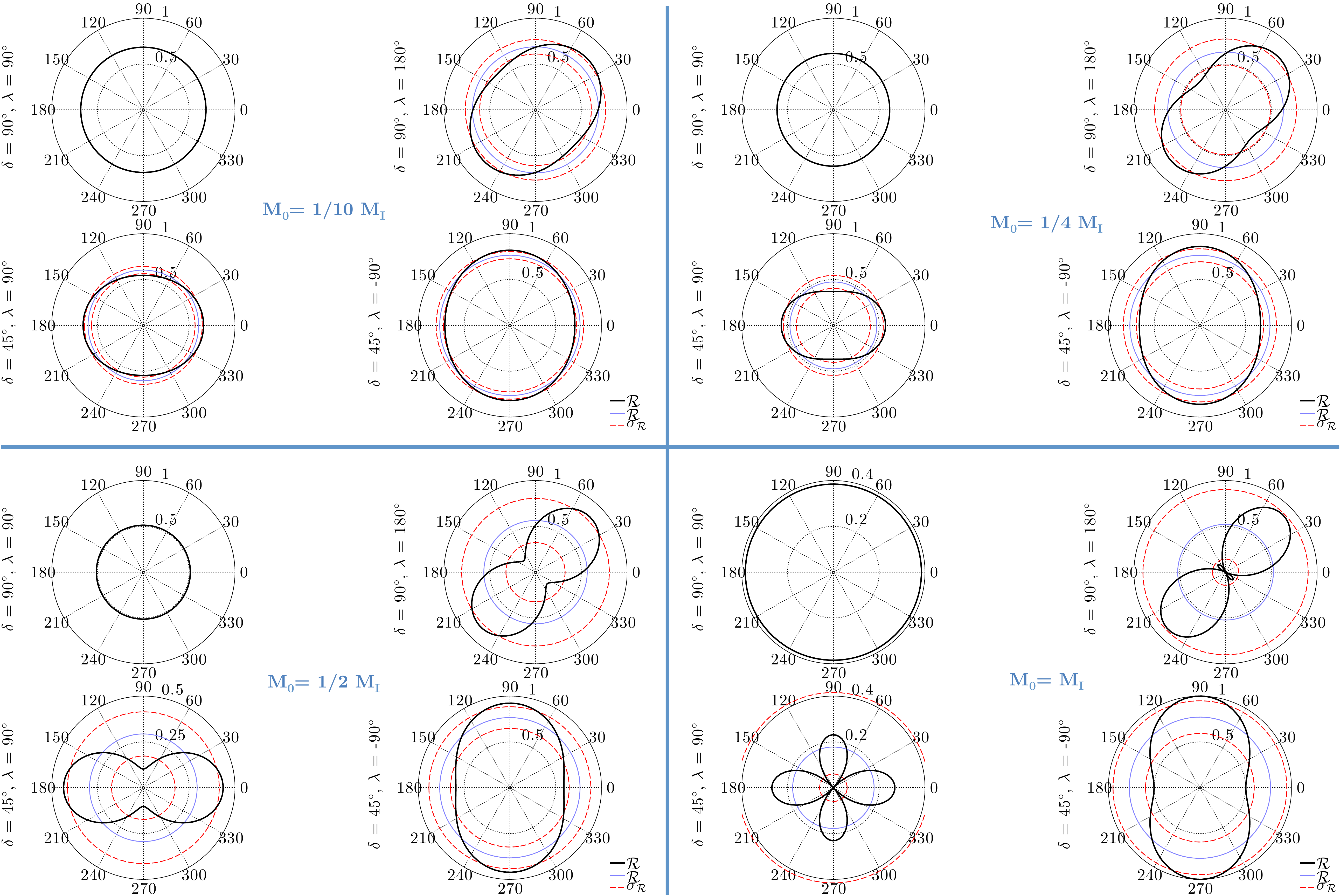}
\captionof{figure}[]
{\narrower Radiation patterns and first two pattern moments for four different faulting geometries of tectonic release and four distinct ratios of isotropic to tectonic release seismic moments ($M_{I}/M_{0}$). Seismic moment ratio panels are separated by thick, blue lines.  Each panel (e.g., top left, $M_{0}$ $=$ $\frac{1}{10}M_{I}$) shows Rayleigh wave radiation patterns (black curves) with their RMS values $\bar{\mathcal{R}}$ (purple curves), superimposed with $\bar{\mathcal{R}}$ $\pm$ $\langle \sigma_{\mathcal{R}} \rangle$ (red, dashed lines). The total seismic moment ($M_{0} + M_{I}$) scales all radiation pattern plots. We emphasize that figure labels $\langle \sigma_{\mathcal{R}} \rangle$ in this figure exclude $\langle \rangle$ (due to lack of visibility), but differs from the variance present in an unknown random process, as considered under our Case II analyses.}
 \label{fig:radPatsFaults}
\end{figure*}
\clearpage
To compute both moments, we assume the source is embedded within a vertically stratified half-space and surrounded by an imaginary, unit radii cylinder to a depth equal to the dominant wavelength $\lambda$ of the radiated elastic energy; the current context provides little risk to confusing wavelength with screening parameter $\Lambda$ and we therefore reuse this conventional symbol. Regardless of symbolism, we then integrate the squared radiation pattern $\mathcal{R}^{2}$ over our cylinder, and normalize the integrand over the cylinder surface area. This process is analogous to using Gaussian surfaces in electrostatics to compute electric fields \cite[Chapter 2]{Griffiths20171}. In our case, this integrated radiation pattern is (from eq. \ref{eq:tecRadPat}):
\begin{equation}
\label{eq:rbarInt}
\begin{split}
\sqrt{\langle \mathcal{R}^{2} \rangle} &= \sqrt{ \bar{\mathcal{R}}^{2} + \langle \sigma_{\mathcal{R}}^{2} } \rangle \\
&= \cfrac{  \sqrt{ \lambda \int_{0}^{2 \pi} d \psi  \left[\, \bar{\mathcal{R}}  + DS \cos( 2 \psi )  +  SS \sin( 2 \psi )\,\right]^{2} } } { \sqrt{2 \pi \lambda } } \\
\end{split}
\end{equation}
The radiation coefficient weights $\left[ 1,\cos (2\psi ), \sin (2\psi )\right]$ in eq. \ref{eq:rbarInt} are mutually orthogonal over the azimuthal interval $[0, 2\pi]$. Hence, cross terms such as $DS \cos (2\psi )\cdot SS \sin (2\psi )$ integrate to zero. The remaining terms reduce eq. \ref{eq:rbarInt} to:
\begin{equation}
\label{eq:rbar}
\sqrt{\langle \mathcal{R}^{2} \rangle} = \sqrt{ \bar{\mathcal{R}} ^{2} + \cfrac{1}{2}\left( DS^{2} + SS^{2} \right) }.
\end{equation}
Comparing to eq. \ref{eq:rbarInt}, we conclude:
\begin{equation}
\label{eq:rvar}
\langle \sigma_{\mathcal{R}}^{2} \rangle = \cfrac{1}{2}\left( DS^{2} + SS^{2} \right).
\end{equation}
The covariance $\langle \sigma_{\mathcal{R}_{i} \mathcal{R}_{j} } \rangle$ $=$ $\text{cov}\left( \mathcal{R}_{i}, \mathcal{R}_{j}\right)$ similarly quantifies how samples of the deterministic radiation pattern correlate at distinct azimuths $\psi_{i}$ and $\psi_{j}$: 
\begin{equation}
\label{eq:covR}
\begin{split}
\langle \sigma_{\mathcal{R}_{i} \mathcal{R}_{j} } \rangle \propto \int_{0}^{2\pi} \int_{0}^{2\pi}& d\psi_{i} d\psi_{j} \left( \mathcal{R}_{i} -\bar{\mathcal{R}}_{i} \right) \left(  \mathcal{R}_{j} - \bar{\mathcal{R}}_{j} \right) 
\\
= \int_{0}^{2\pi} \int_{0}^{2\pi}&  d\psi_{i} d\psi_{j} \left(DS \cos( 2 \psi_{i} )  +  SS \sin( 2 \psi_{i} )\right) \times 
\\
&\left(DS \cos( 2 \psi_{j} )  +  SS \sin( 2 \psi_{j} )\right)
\end{split}
\end{equation}
Products like $\cos( 2 \psi_{i} )\sin( 2 \psi_{i} )$ also integrate to zero. Therefore, for $i \ne j$:
\begin{equation}
\label{eq:covRzero}
\langle \sigma_{\mathcal{R}_{i} \mathcal{R}_{j} } \rangle = 0
\end{equation}
Radiation pattern observations collected from distinct azimuths are therefore statistically independent, when the prior distributions on $\psi_{i}$ and $\psi_{j}$ are uniform. Because this model is a mathematical idealization, we concede that observations collected from sufficiently proximal azimuths are likely to be correlated. We can conclude, however, that the total covariance in an observed radiation field is representable as $\sigma_{M}^{2} \boldsymbol{I}$ when sensor deployments are sufficiently sparse. Fig. \ref{fig:radPatsFaults} illustrates relationships between deterministic radiation pattern moments for several fault types.
\section{Derivation of the Case I Screening Statistic} \label{app:CaseI}
\setcounter{equation}{0}
We apply the hypothesis test of eq. \ref{eq:HypotTest-vec} and combine eq. \ref{eq:altPdf} through eq. \ref{eq:maxLikeGLR} to compute an equivalent GLRT $L_{\boldsymbol{\theta}}^{(2)}\left( \boldsymbol{\mathcal{R}} \right)$:
\begin{equation}
L_{\boldsymbol{\theta}}^{(2)}\left( \boldsymbol{\mathcal{R}} \right)  \triangleq (N-3) \cdot \frac{ \vert \vert  \boldsymbol{\mathcal{R}} - \boldsymbol{H} \hat{\boldsymbol{\theta}}_{0}  \vert \vert^{2} }{  \vert \vert  \boldsymbol{\mathcal{R}} - \boldsymbol{H} \hat{\boldsymbol{\theta}}_{1}  \vert \vert^{2} } 
\label{eq:GLRRatio}
\end{equation}
The MLEs for the unknown parameters are \cite[pg. 252, eq. 8.52]{Kay19931}:
\begin{equation}
\begin{split}
\hat{\boldsymbol{\theta}}_{1} &=  [ \boldsymbol{H}^{\text{T}} \boldsymbol{H} ]^{-1}\boldsymbol{H}^{\text{T}} \boldsymbol{\mathcal{R}} 
\\
\hat{\boldsymbol{\theta}}_{0} &= \hat{\boldsymbol{\theta}}_{1}  -  [ \boldsymbol{H}^{\text{T}} \boldsymbol{H} ]^{-1} \boldsymbol{A}^{\text{T}} \left[ \boldsymbol{A} \left[ \boldsymbol{H}^{\text{T}} \boldsymbol{H} \right]^{-1} \boldsymbol{A}^{\text{T}} \right]^{-1} \boldsymbol{A} \hat{\boldsymbol{\theta}}_{1} 
\\
\hat{\sigma}_{M \,i}^{2} &= \max\left\{ \cfrac{\vert \vert  \boldsymbol{\mathcal{R}} - \boldsymbol{H} \hat{\boldsymbol{\theta}}_{i}  \vert \vert^{2} }{N} - \sigma_{\mathcal{R}}^{2}, \,\,0 \right\}
\end{split}
\label{eq:defineTheta}
\end{equation}
in which $i$ $=$ $0,1$ indicates distinct noise variance estimates under hypothesis $i$, and $\sigma_{\mathcal{R}}^{2}$ $=$ $0$ when $i$ $=$ $0$. To reduce $L_{\boldsymbol{\theta}}^{(2)}\left( \boldsymbol{\mathcal{R}} \right)$ into a more interpretable test statistic, we rewrite the numerator and denominator of eq. \ref{eq:GLRRatio} using projector matrices. These matrices are documented in eq. \ref{eq:projMatrix} and the text that immediately follows eq. \ref{eq:projMatrix}. With these definitions, the Case I test statistic is a ratio of two statistically independent quadratic forms:
\begin{equation}
\label{eq:subpsaceStat}
\begin{split}
L_{\boldsymbol{\theta}}^{(2)}\left( \boldsymbol{\mathcal{R}} \right) &=  (N-3) \cdot  \cfrac{ \bigr \Vert \boldsymbol{P}_{\boldsymbol{H}}^{\perp} \boldsymbol{\mathcal{R}} + \boldsymbol{P}_{\boldsymbol{X}}\boldsymbol{\mathcal{R}} \bigr \Vert^{2} } { \bigr \Vert \boldsymbol{P}_{\boldsymbol{H}}^{\perp} \boldsymbol{\mathcal{R}} \bigr \Vert^{2}},
\end{split}
\end{equation}
so that $\boldsymbol{P}_{\boldsymbol{X}}$ projects onto rank-1 subspace $\text{span}\{\boldsymbol{X}\}$. Matrices $\boldsymbol{P}_{\boldsymbol{H}}^{\perp}$ and $\boldsymbol{P}_{\boldsymbol{X}}$ project onto orthogonal spaces since $\boldsymbol{P}_{\boldsymbol{H}}\boldsymbol{X}$ $=$ $\boldsymbol{X}$. The Pythagorean identity therefore implies that $\bigr \Vert \boldsymbol{P}_{\boldsymbol{H}}^{\perp} \boldsymbol{\mathcal{R}} + \boldsymbol{P}_{\boldsymbol{X}}\boldsymbol{\mathcal{R}} \bigr \Vert^{2}$ $=$ $\bigr \Vert \boldsymbol{P}_{\boldsymbol{H}}^{\perp} \boldsymbol{\mathcal{R}}\bigr \Vert^{2}$ $+$ $\bigr \Vert \boldsymbol{P}_{\boldsymbol{X}}\boldsymbol{\mathcal{R}} \bigr \Vert^{2}$.  This factorization simplifies the screening statistic into a reduced form $L_{\boldsymbol{\theta}}\left( \boldsymbol{\mathcal{R}} \right)$ that is ratio of two statistically independent terms. We then scale $L_{\boldsymbol{\theta}}\left( \boldsymbol{\mathcal{R}} \right)$ by the ratio of the alternative hypothesis to null hypothesis signal variance $(\sigma_{M}^{2} + \sigma_{\mathcal{R}}^{2}) / \sigma_{M}^{2}$:
\begin{equation}
L_{\boldsymbol{\theta}}\left( \boldsymbol{\mathcal{R}} \right) = (N-3) \cfrac{\sigma_{M}^{2}}{\sigma_{M}^{2} + \sigma_{\mathcal{R}}^{2}} \cdot \cfrac{ \bigr \Vert \boldsymbol{P}_{\boldsymbol{X}}\boldsymbol{\mathcal{R}} \bigr \Vert^{2} } { \bigr \Vert \boldsymbol{P}_{\boldsymbol{H}}^{\perp} \boldsymbol{\mathcal{R}} \bigr \Vert^{2}} 
\label{eq:testStat1red}
\end{equation}
The test statistic $L_{\boldsymbol{\theta}}\left( \boldsymbol{\mathcal{R}} \right)$ has noncentral-$F$ distribution parameterized by scalar $\Lambda$ when $\mathcal{R}_{k}$ includes additive Gaussian noise ($1 \le k \le N$). We write its distributional dependence as $L_{\boldsymbol{\theta}}\left( \boldsymbol{\mathcal{R}} \right)$ $\sim$  $\mathcal{F}_{1,N-3}\left( \Lambda>0 \right)$ and the cumulative, noncentral $\mathcal{F}$ distribution for $L_{\boldsymbol{\theta}}\left( \boldsymbol{\mathcal{R}} \right)$ as eq. \ref{eq:Ldistrib}. Scalar $\Lambda$ has the general quadratic form when $\boldsymbol{A}\boldsymbol{\theta}$  $=$ $\boldsymbol{b}$ under $\mathcal{H}_{0}$:
\begin{equation}
\label{eq:genFormNonCentral}
\begin{split}
\Lambda &= \cfrac{ \left( \boldsymbol{A}\boldsymbol{\theta}_{1}  - \boldsymbol{b}  \right)^{\text{T}} \left[ \boldsymbol{A} \left( \boldsymbol{H}^{\text{T}} \boldsymbol{H} \right)^{-1} \boldsymbol{A}^{\text{T}} \right]^{-1} \left( \boldsymbol{A}\boldsymbol{\theta}_{1}  - \boldsymbol{b}  \right)} { \sigma_{M}^{2} + \sigma_{\mathcal{R}}^{2} }
\end{split}
\end{equation}
in which $\boldsymbol{\theta}_{1}$ is the parameter vector under $\mathcal{H}_{1}$. Applied to the current problem in which $\boldsymbol{b}$ $=$ $0$ and eq. \ref{eq:systemMatrix} defines $\boldsymbol{\theta}_{1}$ $=$ $[ \Delta  \bar{\mathcal{R}},DS,SS ]^{\text{T}}$, this scalar becomes:
\begin{equation}
\label{eq:noncentral}
\begin{split}
\Lambda &= \cfrac{ ( DS + SS) \left[ \boldsymbol{A} \left[ \boldsymbol{H}^{\text{T}} \boldsymbol{H} \right]^{-1} \boldsymbol{A}^{\text{T}} \right]^{-1} ( DS + SS) }{ \sigma_{M}^{2} + \sigma_{\mathcal{R}}^{2} }
\end{split}
\end{equation}
The scalar $\Lambda$ completely quantifies the screening capability of $L_{\boldsymbol{\theta}}\left( \boldsymbol{\mathcal{R}} \right)$ to test between $\mathcal{H}_{0}$ and $\mathcal{H}_{1}$. Specifically, increasing values of $\Lambda$ decrease the overlap between the density functions for test statistics that describe cylindrically symmetric ($\Lambda$ $=$ $0$) and faulting sources ($\Lambda$ $>$ $0$). To factor $\Lambda$ into its source-dependent and deployment depend parts, we distribute the scalar $\boldsymbol{A} \left[ \boldsymbol{H}^{\text{T}} \boldsymbol{H} \right]^{-1} \boldsymbol{A}^{\text{T}} $ from $(DS + SS)^{2}$. This factorization results in eq. \ref{eq:factorLambda}.
\section{The Density function for Case II} \label{sec:CaseIIpdf}
\setcounter{equation}{0}
This appendix details the development and performance of the Case II test statistic $L_{\boldsymbol{\theta}}\left( \boldsymbol{\mathcal{R}} \right)$ (we redundantly label the Case I statistic with the same symbol). We first note that the multi-valued inverse of $x(s)$ that we write as $s(x)$ in the PDF arguments of eq. \ref{eq:Zstat} is not a proper function. Rather, $x(s)$ $=$ $s - N \ln( s )$ is minimized at $s$ $=$ $N$, monotonically decreases for all $s$ $<$ $N$, and increases for $s$ $>$ $N$. The Lambert function $W(\bullet)$ compactly represents this multi-valued inverse $s(x)$:
\begin{equation}
\label{eq:LambertFun}
s(x) = -N \,W\left( - \cfrac{e^{-\frac{x}{N}}}{N} \right), \,\, \text{where:} \,\, x > N - N \ln(N)
\end{equation}
in which the properties of $W(\bullet)$ constrain $s(x)$ to be non-negative. To implement the change-of-variables in eq. \ref{eq:Zstat}, we write the PDF $f_{X}\left(x ; \mathcal{H}_{i} \right)$ as a two-term sum over each single valued branch of the Lambert function ($i=0,1$):
\begin{equation}
\label{eq:CaseIIPdf}
\begin{split}
f_{X}\left(x ; \mathcal{H}_{i} \right) = f_{\chi_{N-3}^{2}}\left(  s_{-1}(x) \right) \biggr \vert \cfrac{ds_{-1}}{dx} \biggr \vert + f_{\chi_{N-3}^{2}}\left(  s_{0}(x) \right) \biggr \vert \cfrac{ds_{0}}{dx} \biggr \vert
\end{split}
\end{equation}
in which $s_{-1}(x)$ is the upper branch of the function in eq. \ref{eq:LambertFun} (Fig. \ref{fig:Lamberts}(a), solid curve) and $s_{0}(x)$ is the lower branch of the function in eq. \ref{eq:LambertFun} (Fig. \ref{fig:Lamberts}(a), dashed curve). We write the two derivatives as (algebra omitted):
\begin{equation}
\label{eq:CaseIIPdfDeriv}
\begin{split}
\cfrac{ds_{k}}{dx} &= \cfrac{W_{k} \left( \cfrac{e^{-\frac{x}{N}}}{N} \right) }{W_{k} \left( \cfrac{e^{-\frac{x}{N}}}{N} \right) + 1} 
\end{split}
\end{equation}
in which $k=-1,0$. The functional form of the PDF for chi-square statistic $f_{\chi_{N-3}^{2}}\left( \bullet \right)$ includes a zero noncentrality parameter under both hypotheses on radiation pattern shape. We explain this by expanding the numerator of ratio $\Vert \boldsymbol{P}_{\boldsymbol{H}}^{\perp} \boldsymbol{\mathcal{R}} \bigr \Vert^{2} / \sigma_{M}^2$. First, we note that the radiation pattern $\boldsymbol{\mathcal{R}}$ is a sum of a linear model $\boldsymbol{H} \boldsymbol{\theta}$ and noise. Three of the four terms in this expansion include products of the projector matrix $\boldsymbol{P}_{\boldsymbol{H}}^{\perp}$ and $\boldsymbol{H}$. Each of these terms are zero. This leaves $\boldsymbol{n}^{\text{T}} \boldsymbol{P}_{\boldsymbol{H}}^{\perp} \boldsymbol{n}$, in which $\boldsymbol{n}$ is a vector of zero-mean noise with the same dimensions as $\boldsymbol{\mathcal{R}}$. Such quadratic forms have central $\chi^{2}_{N-3}(0)$ distributions when scaled by $1/\sigma_{M}^{2}$. 

The ratio with quadratic form $\Vert \boldsymbol{P}_{\boldsymbol{X}}\boldsymbol{\mathcal{R}} \bigr \Vert^{2} / \sigma_{M}^{2}$ determines the PDF for $f_{Y}\left(\bullet ; \mathcal{H}_{i} \right)$. In this case, it is identical to the numerator of eq. \ref{eq:testStat1red} and is distributed as $\chi_{1}^{2}(\Lambda)$ when scaled by $1/ \left( \sigma_{M}^{2} + \sigma_{\mathcal{R}}^{2} \right)$. The scalar $\Lambda$ is zero under $\mathcal{H}_{0}$, and equates to eq. \ref{eq:noncentral} under $\mathcal{H}_{1}$. Given each PDF (eq. \ref{eq:CaseIIPdf} and $f_{\chi_{1}^{2}}\left(\bullet \right)$), we compute the convolution of eq. \ref{eq:Zstat} numerically in the Fourier domain with the characteristic function method \cite[eq. 12]{Carmichael20201}:
\begin{equation}
f_{Z} \left(  z   ;  \mathcal{H}_{i}  \right) =  \mathcal{F}^{-1}\left\{  \mathcal{F} \left\{f_{X}\left(x ; \mathcal{H}_{i} \right) \right\} \mathcal{F}\left\{f_{Y}\left(y ; \mathcal{H}_{i} \right) \right\} \right\}.
\label{eq:fZPdf}
\end{equation}
in which $\mathcal{F}$ is the Fourier Transform (FFT) and $\mathcal{F}^{-1}$ is its inverse. The computational grid and resolution of fast-Fourier transforms (\texttt{fft.m} and \texttt{ifft.m} in \texttt{MATLAB}) control the precision of eq. \ref{eq:fZPdf}. Consequently, our estimate $\hat{\eta}$ for a threshold $\eta$ that maintains a constant false attribution probability is necessarily inexact. To compute $\hat{\eta}$, we treat numerically unique values of the null PDF $f_{Z} \left(  z   ;  \mathcal{H}_{0}  \right)$ output by eq. \ref{eq:fZPdf} as an independent variable and their corresponding grid values of $z$ as a dependent variable. We then linearly interpolate $z$ at the grid value for $\text{Pr}_{FA}$ to produce the estimate $\hat{\eta}$ and symbolize these numerical operations for reference:
\begin{equation}
\label{eq:interpCaseIIThr}
\hat{\eta} = \text{interp}\left(f_{Z} \left(  z   ;  \mathcal{H}_{0}  \right)_{l},  z_{l},  \text{Pr}_{FA} \right),
\end{equation}
in which index $l$ indicates a numerically unique grid value. Fig.\ref{fig:Lamberts}(b) illustrates eq. \ref{eq:fZPdf} estimates for PDFs under the competing hypotheses. We selected $\Lambda$ $=$ $10$ under $\mathcal{H}_{1}$ and a false attribution probability of $\text{Pr}_{FA}$ $=$ $5\cdot10^{-3}$. The FFT inversion appears stable when $\Lambda$ is not too large \cite[Section S2]{Carmichael20201}. Sources that trigger radiation patterns with high SNR faulting signals will be obvious to screen, and we therefore do not consider this restriction on $\Lambda$ of practical significance. To compute screening probability values for each of the focal mechanisms shown in Fig. \ref{fig:CaseIIvCaseI}, we integrate the PDFs in eq. \ref{eq:interpCaseIIThr} over a 512 point grid to construct a CDF for each source focal mechanism. We then repeated this integration over a 500 point grid of faulting SNIR values (horizontal axis). We then estimate each screening probability by interpolating the CDF at a screening threshold that is consistent with a false alarm value of $\text{Pr}_{FA}$ $=$ $10^{-3}$.

\setcounter{figure}{0}
\begin{figure*}
\centering
\includegraphics[width=\textwidth]{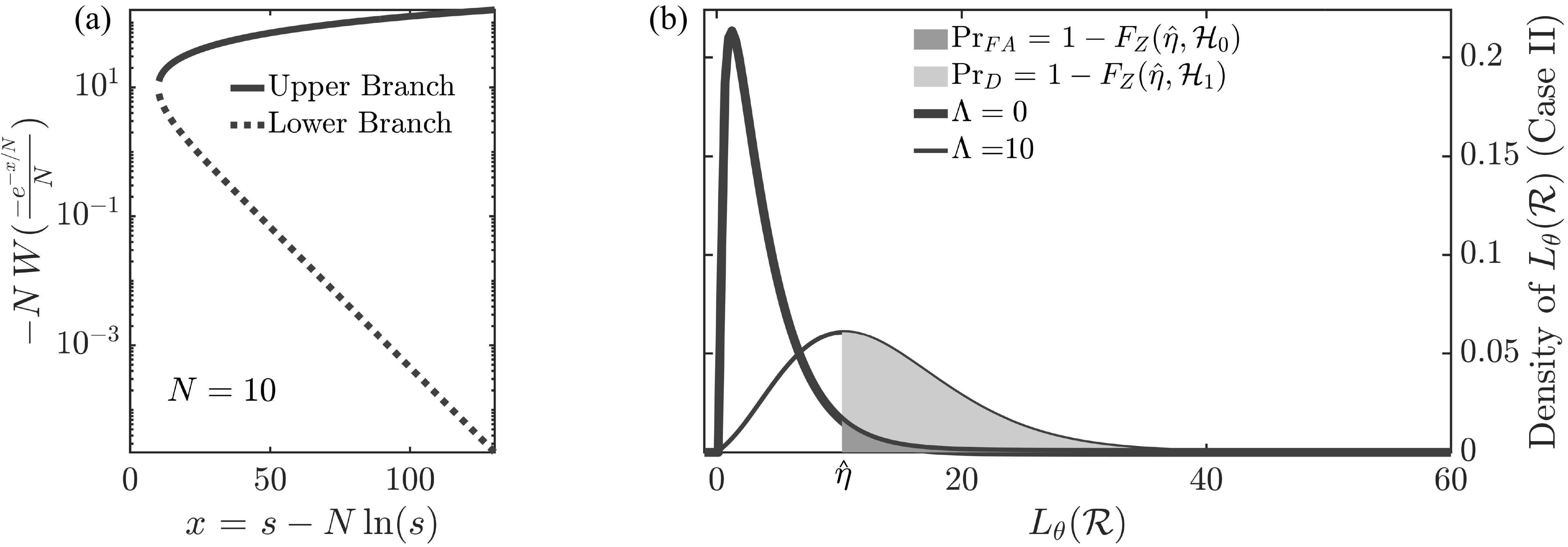}
\caption[]
{\narrower Mathematical and statistical properties of the arguments and outputs of the Case II screening statistic. (\textbf{a}): Both branches of the Lambert-W function. The upper branch $s_{-1}(\bullet)$ and lower branch $s_{0}(\bullet)$ of the function shown in eq. \ref{eq:LambertFun}, with argument $-e^{-\frac{x}{N}} / N$. This example uses $N$ $=$ $12$ sensors. (\textbf{b}): The density function $f_{Z}\left(z; \mathcal{H}_{i}\right)$ for the Case II test statistic under both $\mathcal{H}_{0}$ and $\mathcal{H}_{1}$. The noncentrality parameter $\Lambda$ matches that for Case I.}
 \label{fig:Lamberts}
\end{figure*}
\end{appendices}

\end{document}